\documentclass[acmsmall,screen]{acmart}
\AtBeginDocument{%
  \providecommand\BibTeX{{%
    \normalfont B\kern-0.5em{\scshape i\kern-0.25em b}\kern-0.8em\TeX}}}

\usepackage{caption}
\usepackage[skip=0cm,list=true,labelfont=it]{subcaption}

\setcopyright{acmcopyright}
\copyrightyear{2022}
\acmYear{2022}
\acmDOI{XXXX/XXXXX}

\begin{document}

\title[Heterogeneity in Algorithm-Assisted Decision-Making]{Heterogeneity in Algorithm-Assisted Decision-Making: A Case Study in Child Abuse Hotline Screening} 


\author{Lingwei Cheng}
\affiliation{%
  \institution{Carnegie Mellon University}
  \city{Pittsburgh}
  \country{USA}}
\email{lingweic@andrew.cmu.edu}

\author{Alexandra Chouldechova}
\affiliation{%
  \institution{Carnegie Mellon University}
  \city{Pittsburgh}
  \country{USA}}
\email{achoulde@andrew.cmu.edu}

\begin{abstract}
Algorithmic risk assessment tools are now commonplace in public sector domains such as criminal justice and human services.  These tools are intended to aid decision makers in systematically using rich and complex data captured in administrative systems.  In this study we investigate sources of heterogeneity in the alignment between worker decisions and algorithmic risk scores in the context of a real world child abuse hotline screening use case.  Specifically, we focus on heterogeneity related to worker experience.  We find that senior workers are far more likely to screen in referrals for investigation, even after we control for the observed algorithmic risk score and other case characteristics.  We also observe that the decisions of less-experienced workers are more closely aligned with algorithmic risk scores than those of senior workers who had decision-making experience prior to the tool being introduced. While screening decisions vary across child race, we do not find evidence of racial differences in the relationship between worker experience and screening decisions.  Our findings indicate that it is important for agencies and system designers to consider ways of preserving institutional knowledge when introducing algorithms into high employee turnover settings such as child welfare call screening.  
\end{abstract}

\ccsdesc[500]{Human-centered computing~User studies}
\ccsdesc[500]{Information systems~Decision support systems}
\ccsdesc[500]{Applied computing~Computing in government}

\keywords{human-computer interaction, child welfare, organizational behavior}

\maketitle

\textcolor{red}{Preprint accepted by CSCW 2022}

\section{Introduction}
\label{sec:introduction}
Data-driven algorithmic tools are increasingly being adopted in government and industry to aid decision makers in the systematic use of complex data. The introduction of new technology into the workplace can disrupt and profoundly change workplace dynamics by directly influencing workers' return to expertise \cite{routinebias2014}, incentives to adapt to technology \cite{NBERw27950}, and decision quality \cite{Dawes1989, Dietvorst15}.  A number of recent studies have assessed the impact of algorithmic tools on decision-making in domains such as health care \cite{giannini2019machine, Obermeyer447}, criminal justice \cite{10.1093/qje/qjx032, Kehl2017AlgorithmsIT}, education \cite{10.1145/3303772.3303814}, and human services \cite{Kube_Das_Fowler_2019}.  We contribute to this emerging literature by investigating sources of heterogeneity in the alignment between worker decisions and algorithmic risk scores in the context of child abuse hotline screening.  Specifically, we focus on heterogeneity related to worker experience.  Given the widespread racial disparities in the child welfare system, we also investigate whether the association between algorithm-assisted screening decisions and worker experience varies with the race of children involved in the referral.  

In the context of child welfare services, many states have adopted algorithmic decision support tools to improve the quality and consistency of decisions \cite{AFST_desc, eckerd_connects_2020}. However, while many studies have rigorously evaluated the theoretical and practical performance of the tools, very few have considered the impact of organizational factors such as workforce characteristics and resource constraints in the decision contexts. According to a comprehensive overview of algorithms used within the U.S. Child Welfare System \cite{10.1145/3313831.3376229}, two factors that are frequently ignored are caseworker characteristics and turnover, which are directly associated with placement instability as characterized by a greater number of placements \cite{doi:10.1080/15433714.2013.788953}. Volatility in staffing can lead to a loss of social, financial and knowledge capital, affecting the training, skill, and resources of the current workforce, who are crucial to providing services to the vulnerable child, youth, and families \cite{united2003child, RYAN2006993}. 

Our work attempts to fill some of this gap by investigating whether call worker screening decisions in the presence of an algorithmic risk assessment tool vary with worker experience, and whether there is further variation with child race.  Specifically, we focus on three main research questions: (i) Are call-worker decisions less aligned with the algorithmic score for more senior workers who worked at the call center before the risk assessment tool was introduced? (ii) Is worker experience associated with differences in the ability to compensate for a technical glitch that resulted in real time scores being calculated incorrectly in certain cases; and (iii) Are there racial differences in the workers' decisions when presented with referrals with the same predicted risk, and does this vary by worker experience?  Our investigation of question (ii) builds upon a previous study \cite{10.1145/3313831.3376638} where the authors explored a unique implementation glitch where risk scores shown to the call screeners were miscalculated.  Our analysis is based on data collected following the deployment of the Allegheny Family Screening Tool (AFST), which was made available by the Allegheny County Department of Human Services in Allegheny County, Pennsylvania, USA.


We show that senior call screeners persistently screen in referrals at higher rates compared to junior and temp screeners when facing referrals with the same characteristics and shown algorithmic scores. 
Similar to the findings of \citet{10.1145/3313831.3376638}, when the shown score is an incorrect estimation of the assessed score by the algorithm, screeners overall are more likely to screen in a referral if the shown score is an underestimation of the assessed score, suggesting they are able to effectively rely on other information and compensate for the miscalculated algorithmic risk score. We do not find evidence that the ability to compensate for these errors is associated with worker experience.    

Overall, we find that, while there are differences in screen-in rates between referrals on black children and white children at most values of the risk score, those differences are attributable to other explanatory factors.  The racial gap is attributable to the association between race, and exposure to the implementation glitch as well as other contextual characteristics.  This is a notable finding given significant concerns regarding the over-screening of racial minorities \cite{_rauktis_role_2010}.

Our study fills an important gap in the evaluation of algorithmic assessments in child welfare systems by addressing the relationship between decision-makers' level of experience outside of algorithm-assisted settings and their use of algorithmic predictions. While prior literature evaluates screening decisions at an aggregate level without controlling for potential explanatory variables and inferring different stories, we highlight other mechanisms through which one could observe racial gaps in screening decisions. Given that maintaining a stable workforce has been a major challenge of child welfare agencies, this case study may be helpful to agencies in developing guidelines on training and maintaining institutional knowledge in algorithm-assisted decision-making settings.

Beyond the specific domain of child welfare, our empirical investigation on user heterogeneity contributes to our understanding of reliance on automation, particularly on the role of user characteristics in mediating reliance \cite{Goddard11, Lee04, Veale2018}. Our results regarding race-related differences also contribute to the growing literature on decomposing algorithmic bias and disparate interactions with algorithms \cite{benyin20}. Both are important in ensuring effective and fair human-AI collaboration.

\section{Related Work}
\label{sec:related_work}
Developing a better understanding of algorithm-assisted decision-making is a core thrust of research on computer-supported collaborative work.  A significant body of work has explored factors that influence the degree to which users trust AI-based algorithmic tools across different domains (see e.g., \cite{glikson2020human} for a recent survey).   Other work has examined the role of transparency, explainability, and interpretability \cite{schmidt2020transparency, poursabzi2021manipulating}, as well as stated and apparent accuracy \cite{yin2019understanding} on the uptake of algorithmic predictions and recommendations.  The work of~\citet{Dietvorst15} documented the phenomenon of algorithm aversion, wherein users were found to be less willing to rely on algorithmic predictions after seeing them err, observing in a later study that reliance improves if humans are able to alter the predictions to even a small degree~\cite{dietvorst2016overcoming}.

Empirically, it has been shown both in lab and real-life settings that people alter their decision-making in the presence of an algorithmic decision support tool, and they can interact with it in a way that produces disparate outcomes. According to a controlled experiment on Amazon's Mechanical Turk \cite{benyin20} using a pretrial release algorithm, participants presented with a risk assessment made predictions that were less accurate than the risk assessment’s, and could not effectively evaluate the accuracy of their own or the risk assessment’s predictions. Additionally, they exhibited behaviors fraught with disparate interactions by assigning higher risk prediction to black defendants than to white ones. This pattern of disparate interaction is due to participants' increasing response to risk scores given a particular race, and features that are unevenly distributed across race. Another experiment based on hypothetical online hiring setting similarly showed that fair ranking algorithms can help increase the number of underrepresented candidates selected but only when the minority candidates share similar features as the majority group and are not subject to employers' persistent gender bias \cite{suhr2021does}. However, it also found that participants who assumed the roles of employers are cognizant of the unfairness and underrepresentation in the candidate pool and are capable of reinforcing their notion of fairness. Nevertheless, to extrapolate these findings from crowdworker studies to real world settings, one needs to further consider experiment participants' lack of context and experience compared to actual decision-makers.

In real-world policy settings, there have been many examples in criminal justice \cite{stanford16}, education \cite{doi:10.1177/0956797615570365}, child welfare \cite{doi:10.1086/675373}, and medical research \cite{babitsch} suggesting that judgement is closely associated with the characteristics of the decision-makers themselves. Those characteristics can include their own demographics and experiences, as well as more subtle cultural norms and beliefs. 

In the area of criminal justice, a study on algorithmic bail recommendation in Kentucky found that racial disparities for low- and moderate-risk defendants increased even though the policy recommended the same treatment for both black and white groups with those risk levels \cite{albright2019if}. Judges in whiter counties were more receptive the algorithm's default recommendation to release low-risk individuals than judges in blacker counties. Judges also more often overrode the recommended default of non-financial bonds for moderate-risk black defendants than similar moderate-risk white defendants. Providing judges with risk assessment about a defendant was also found to increase the severity of their sentences for relatively poor—but not affluent—defendants \cite{skeem2020impact}. Thus, there can be great heterogeneity in the impact of a decision support tool across geographic locations and types of decision-makers even though the same tool is provided. At the same time, objectives, incentives, and interpretations of information could vary widely across domains. 

In child welfare systems, stakeholders including families, providers and specialists are commonly concerned with potential bias that is inherent in the data, algorithm, and on the part of caseworkers involvement in the decision processes \cite{10.1145/3290605.3300271}. Tensions between algorithm and caseworkers could arise from a lack of alignment of resources and incentive structures as well as from ideological mismatch \cite{10.1145/3406865.3418331}. Caseworkers can diverge from a decision support tool's recommendation because they assess the meaning of demographic factors and caregivers' histories differently \cite{Bosk18}. In a comparison between a multidisciplinary team decision-making model and a clinically-based decision support algorithm, the concordance rate of their decisions can vary depending on the restrictiveness of the placement setting and age of children \cite{CHOR2013871}. Given the concerns and tensions within the child welfare systems, technical solutions are seen as insufficient for ensuring the system is perceived as fair and just. 

With regard to the AFST, a previous study has shown that child maltreatment call screeners are able to mitigate the effect of erroneous risk assessment scores by adhering less to the algorithm's risk assessment when the score displayed was an incorrect estimate of risk \cite{10.1145/3313831.3376638}. Our paper differs from this earlier work in several key ways.  First, our questions examine heterogeneity in algorithm-assisted decision-making across screeners' experience level, which was not considered in the earlier study.  Second, our methodological approach is different.  Instead of relying primarily on descriptive statistics, we use regression models that account for both correlations in the screeners' decisions and observed explanatory/confounding factors such as time, supervisor effects and case characteristics.  Thus while our study context is similar, our primary questions and methodology are different.  



\section{Background and context}
\label{sec:background}
On August 1$^{st}$ 2016, the Allegheny County Health and Human Services Department implemented the Allegheny Family Screening Tool (AFST) as a decision-support tool, with the goal of improving both the accuracy and consistency of decisions made about referrals to the child maltreatment hotline \cite{allegheny_eval}. AFST is a predictive risk model built and trained using data from the county's integrated data system including demographics, justice system involvement, public welfare histories, and behavioral health histories of child victims, parents, and alleged perpetrators. 

\subsection{Worker Training}
\label{sec:background_training}
Prior to implementation, staff were introduced to the tool through a three-hour training session provided to all full-time and occasional call screening staff, intake administrators, and key child welfare administrators. The training provided a brief overview of Predictive Risk Modelling and the application of it within Allegheny County to give participants an understanding of what risk modeling is, how the model was built, and the predictive power of the model. 

\subsection{Call Screener Work Flow}
\label{sec:background_work_flow}
At time of receiving a call, the call screener learns about contextual information from the referral call. They also have access to ClientView, which is a front-end tool that provides screeners access to an integrated data system containing information on all children and adults associated to the referral. After they enter and submit all the information on the current referral, they receive a risk score produced by AFST. 

The AFST score is based on the maximum score (either re-referral or out-of-home placement) across all children associated with the referral at the time of the screening call. The score ranges from 1 to 20 (where 20 is the highest risk and 1 is the lowest). During the period spanned by our data, there was a ``mandatory'' screening protocol in place.  Under this protocol, when the highest out-of-home placement risk score is 18 or higher, the referral is considered a mandatory screen-in that can only be overridden by a supervisor. 

After weighing all the information, the call screener subsequently decides to recommend in their report: 1) Screen-out the referral without, any further evaluation or assessment, 2) Field screen the referral, to assess whether an investigation is warranted, or 3) Screen-in the referral, which is synonymous with conducting a formal investigation.  Answering a call takes between 25 to 35 minutes on average, from collecting all the information on a referral to completing the call report for the supervisor.  

\subsection{Issues affecting deployment}
\label{sec:background_deployment_issues}
Since AFST was first introduced, there are two additional changes in the implementation. First, in the initial months of deployment between August and November 2016, the AFST was not able to produce a risk score for a referral if there was no previous information on the children in the database. This issue was resolved on December 1$^{st}$ 2016. We drop data in the initial implementation period in the subsequent analysis.

Secondly, a glitch in the system led to certain model inputs not being calculated correctly in real time until February 2019. The glitch affected a range of features and commonly resulted in non-zero count and indicator features incorrectly taking on the value $0$.
This in turn produced a discrepancy between the scores shown to the call screeners and the scores that would have been generated correctly. Call screeners were not aware of the issue and only had access to the scores that were shown. They can, however, access the databases where these features are correctly recorded. For analyzing call screeners' behaviors, we use the shown scores since they were the scores call screeners had access to in real time. 

\begin{figure}[h]
\centering
\includegraphics[scale=0.4]{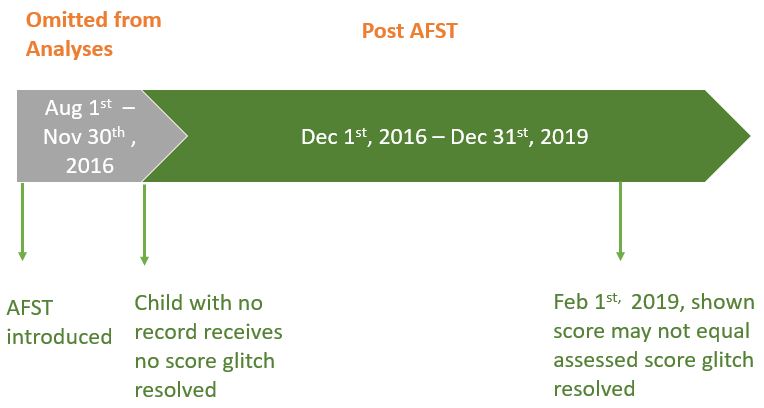}
\caption{Key Implementation Changes}
\label{fig:implementation_changes}
\end{figure}

\section{Data and Descriptive Analyses}
\label{sec:data_desc}
The data used in our study consists of referrals made to the Allegheny County Office of Children, Youth and Families (CYF) between December 1$^{st}$ 2016 and December 31$^{st}$ 2019.  In a number of cases, call screeners and supervisors have no discretion in how to respond to the call.  For instance, while screeners in Pennsylvania have discretion in General Protective Services referrals, which concern suspected neglect or other conditions short of abuse that the reporter believes warrant intervention, they are required to screen in referrals classified as Child Protective Services (allegations of abuse).  Screeners also have no discretion in referrals pertaining to families under active investigation or referrals associated with truancy. Since we are interested in examining worker decision-making, we exclude from analysis all referrals in which screeners and supervisors had no discretion. We also exclude a very small number of referrals whose outcomes are missing.

Each referral can be associated with multiple children, which raises questions surrounding the unit of analysis.  Since decisions are made at the referral level, and screeners see a single score for the referral rather than separate scores for each individual child, we take the referral to be the unit of analysis for the purpose of our study. Note that although certain children in the referral might have low risk scores, call screeners only observe the highest risk score among all children. 

For each referral, we observe key information from Allegheny County's child welfare case management system (KIDS). KIDS is one of the sources of data AFST uses to generate risk scores. It contains demographics such as the age, gender, race, and ethnicity of the victims and perpetrators, counts of allegations, specific types of alleged abuse, and whether the referral was made by a mandatory reporter. It also contains administrative details on the referrals including the timing of the referral, subsequent decisions on screening, field investigation and services provision, as well as the unique identifiers of the screener and supervisor responsible for the decisions.  We use a black-prioritized coding for race, which means that if there is at least one black child on the referral, the race of the referral is coded as black. 

The terminology and notation we use throughout the paper are detailed in Table ~\ref{table:notation}.



\begin{table}[h]
\caption{Terminology and Notation}
\scalebox{0.85}{
\begin{tabular}{@{}lp{3cm}p{8cm}@{}}
\toprule
Term                        & Notation   & Description                                                                                                                                                                                                     \\ \midrule
\textbf{Dependent Variable} &            &                                                                                                                                                                                                                 \\
Screened In                 & Y $\in$ \{0,1\}          & 1 if a referral was screened-in, 0 otherwise                                                                                                                                                                       \\
\textbf{Controls}           &            &                                                                                                                                                                                                                 \\
Assessed Score              & $S \in \{1,\ldots, 20\}$          & The score assessed by the predictive model using the correct features. Maximum over re-replacement and re-referral risks over all children.                                                                     \\
Shown Score                 & $\tilde{S} \in \{1,\ldots,20\}$       & The score shown to the call screeners. It should correspond to the assessed score but is subject to a glitch in the system                                                                                      \\
Black                & black $\in$ \{0, 1\}& 1 if at least one child on the referral was black, 0 otherwise                                                           \\
Screener Type                 & screenerType  $\in$ \{senior, junior, temp\} & A categorical variable indicating if a worker is senior (worked both before and after AFST and at least 300 cases), junor (only started working after AFST and worked at least 300 cases), or temp otherwise                       
    \\
Glitch Type                 & glitchType $\in$ \{over, correct, under\}& A categorical variable indicating if shown score is an over/correct/under-estimation of the assessed score                 
\\
Case Characteristics        & X          & A vector of variables on the characteristics of a referral including count, age, and gender of the victims and perpetrators, total and unique counts of allegations, specific types of alleged abuses (out of the 237 categories), and if the referral was made by mandatory reporters\\
Monthly Fixed Effects       & t          & Monthly fixed effects controlling for the seasonal variation (school year and holidays) in screen-in rate                                                                                                                                 \\
Supervisor Fixed Effects    & j          & Supervisor fixed effects controlling for the variation in supervisors' preference in screen-in decisions               \\            \textbf{Notation for Data Description}           &            &                                                                                \\ 
Shown Mandatory Screen-in   & $\tilde{M}$ $\in$ \{1,0\}          & 1 if the highest out-of-home placement shown score generated in real time is equal to or greater than 18. Supervisor approval is required for screen out.\\ 
\bottomrule
\end{tabular}
}
\label{table:notation}
\end{table}

\subsection{Shown and Assessed Scores}
\label{sec:data_desc_scores}
Our data contains two risk scores for each referral: the \textit{shown} risk score, and what \cite{10.1145/3313831.3376638} termed the \textit{assessed} risk score.    The shown scores refer to scores that call screeners saw in real time. The assessed scores refer to scores that were retrospectively produced using corrected model input features after the glitch was detected and corrected.  Due to the glitch, the shown scores in many instances do not coincide with the assessed scores.  We observe both shown and assessed scores between December 1$^{st}$ 2016 and July 31$^{st}$ 2018.  
For analyzing screening and overriding decisions in connection to the feature computation glitch, we use referrals where both scores are available. 

\begin{figure}[h]
\centering
\includegraphics[scale=0.21]{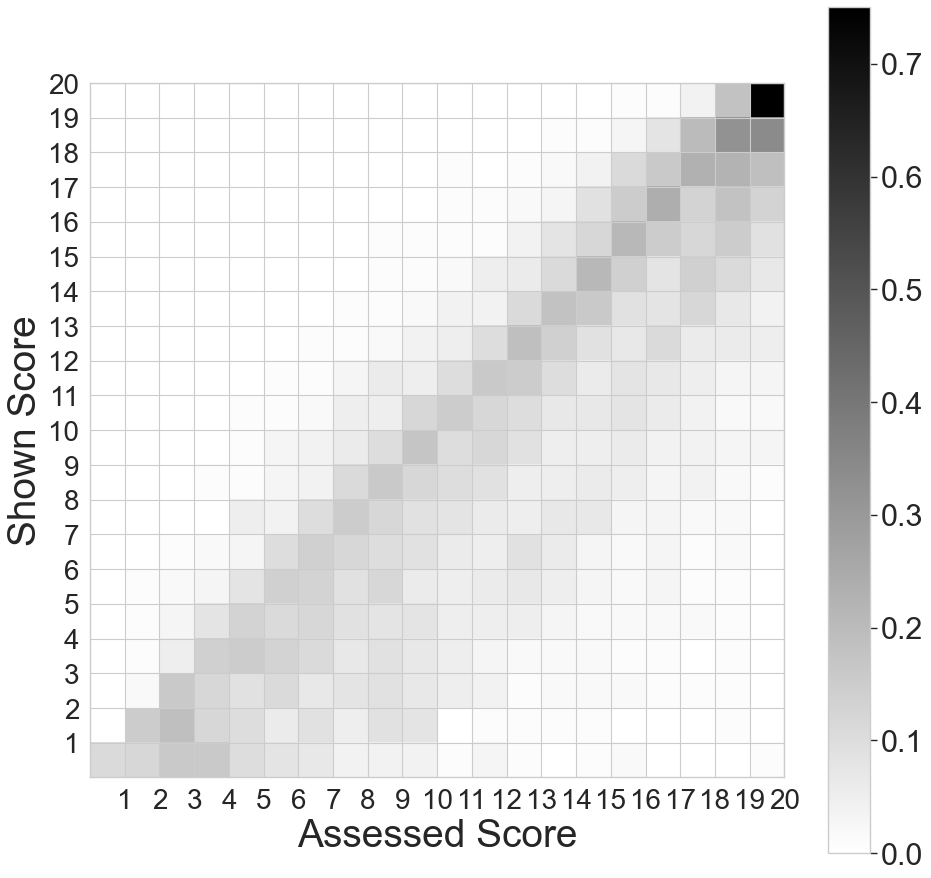}
\caption{Shown vs Assessed Scores. The heatmap shows the density of assessed score conditional on shown score. Shown scores largely overlap with assessed scores, but tend to be an underestimate of the assessed score.}
\label{fig:shown_vs_assessed}
\end{figure}

Figure~\ref{fig:shown_vs_assessed} shows a plot of shown vs. assessed score.  Consistent with the finding in \cite{10.1145/3313831.3376638}, we find that shown scores largely overlap with and tend to be an underestimation of the assessed scores.


\subsection{Screener Type}
\label{sec:data_desc_screener}
The data enables us to define different categories of workers based on the number of calls they screened before and after the deployment of the AFST.  We will use the terms senior, junior, and temp to refer to screeners who worked both  pre/post-AFST, only post-AFST, and on temporary basis, respectively.  A screener is considered to be \textit{senior} if they worked consistently over time on at least 300 referrals pre-AFST and 300 referrals post-AFST. A screener is considered to be \textit{junior} if they worked consistently, their first day of work was after the AFST was deployed, and they worked on at least 300 referrals.  All other workers are coded as \textit{temp}.  The temp category reflects the CYF practice of hiring temp workers to help with case loads throughout the year. Using these definition, we obtain the same number of screeners as reported in Allegheny County's commissioned evaluation study of the AFST implementation \cite{goldhaber-fiebert_prince_2019}. 

As shown in Table ~\ref{table:worker_summary}, there are in total 76 call screeners, with 11 senior screeners, 16 junior screeners, and 49 temp screeners during our period of analysis. Although there are many more temp screeners, they only cover 25.2\% of all referrals. Full-time screeners take on the majority of the workload, with senior screeners processing 40.7\% of all referrals and junior screeners processing 34.1\%. 

\begin{table}[h]\small
\caption{Summary statistics of the caseload across different screener experience types.  The table shows counts and means (M) and standard deviations (SD) for the shown and assessed scores.  The means and standard deviatons are very similar across screener groups.  This indicates that, to the extent that the AFST is an accurate reflection of risk, the different screener types were exposed to similarly risky case mixes. }
\begin{tabular}{@{}cp{1.5cm}p{1.5cm}p{1.5cm}cccccc@{}}
\toprule
       & Number of Screeners & Number of Cases & Percentage of Cases & \multicolumn{2}{p{2.1cm}}{Shown Score (12/1/16-12/21/19)} & \multicolumn{2}{p{2cm}}{Shown Score (12/1/16-7/31/18)}& \multicolumn{2}{p{2cm}}{Assessed Score (12/1/16-7/31/18)} \\ \midrule
       &                     &                 &                     & M               & SD      & M               & SD       & M                & SD              \\
Senior & 11                  & 11,680          & 40.7                & 13.3            & 4.8   & 13.5            & 4.9         & 14.6             & 4.4             \\
Junior & 16                  & 9,803            & 34.1                & 13.0            & 4.8       & 13.5            & 5.0        & 14.3             & 4.6             \\
Temp   & 49                  & 7,253           & 25.2                & 13.0            & 5.0       & 13.1            & 5.0       & 14.2             & 4.7             \\ \bottomrule
\end{tabular}
\label{table:worker_summary}
\end{table}

\subsubsection{Case mixture across worker types}
\label{sec:data_desc_case_mix}
 Table~\ref{table:worker_summary} provides summary statistics of the caseload for each screener type.  We see that the mean shown scores are 13.3 for senior screeners, 13 for junior screeners, and 13 for temp screeners. The mean assessed scores are $14.6$, $14.3$ and $14.2$, for these groups, respectively.  To the extent that the assessed AFST is an accurate reflection of risk, these results suggest that the different screener groups handled approximately equally risky mixes of cases.  We note that in conversations with CYF staff we heard supervisors remark that they do at times assign more complex cases to more experienced screeners.  To the extent that this is the case, the notion of complexity used in making those case assignment decisions does not appear to be captured by shown or assessed AFST scores.
For the purpose of our analysis, it suffices to observe that it is \textit{not} the case that only senior workers can work on high-risk cases. This mitigates the concern that there could be large systematic differences in how calls were processed. The conclusion holds when we look at the subset of referrals between December 1$^{st}$ 2016 and July 31$^{st}$ 2018 when both shown and assessed scores are available, as illustrated in Appendix ~\ref{section:appendix_summary_graphs_risk_distribution}.

\begin{figure}[h]
\centering
\includegraphics[scale=0.25]{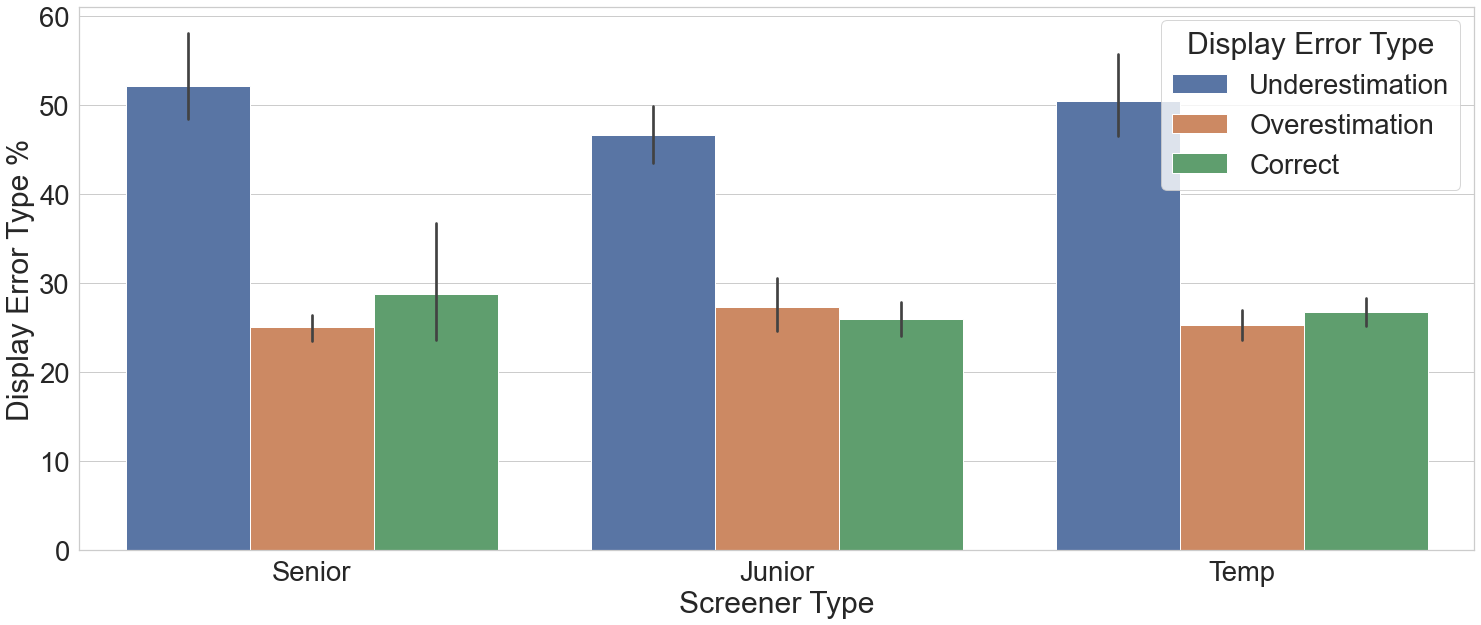}
\caption{Display Error Distribution By Screener Type. Screeners were exposed to similar distributions of underestimated (shown $<$ assessed), overestimated (shown $>$ assessed), and correct (shown $=$ assessed) AFST scores.  Error bars correspond to 95\% confidence intervals.}
\label{fig:display_error_by_screener}
\end{figure}

Figure~\ref{fig:display_error_by_screener} shows the fraction of cases in which the different screener types saw underestimated (shown $<$ assessed), overestimated (shown $>$ assessed), and correct (shown $=$ assessed) AFST scores.  From this plot we see that
screeners were similarly exposed to the glitch in displays of assessed scores. Since the shown score was most often an underestimate of the assessed score, screeners in all groups were most likely to encounter underestimated scores. While there is some variation across screener types on average, many of the confidence intervals are wide and the observed differences are in general not statistically significant. 

\subsubsection{Screening decisions across AFST scores.}
\label{sec:data_desc_screen_in_rate}
\begin{figure}[h]
  \begin{minipage}[t]{0.45\textwidth}
    \includegraphics[width=\textwidth]{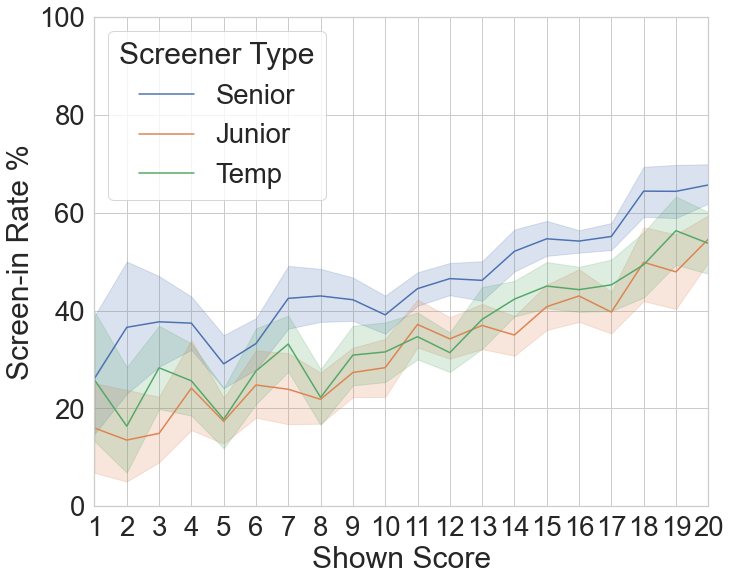}
    \caption{Screen-in Rate by Screener Type. Senior screeners consistently screen in at high rates than junior and temp screeners, conditioned on the same shown score}
    \label{fig:screen_in_by_screener}
  \end{minipage}
  \hfill
  \begin{minipage}[t]{0.45\textwidth}
   \centering
    \includegraphics[width=\textwidth]{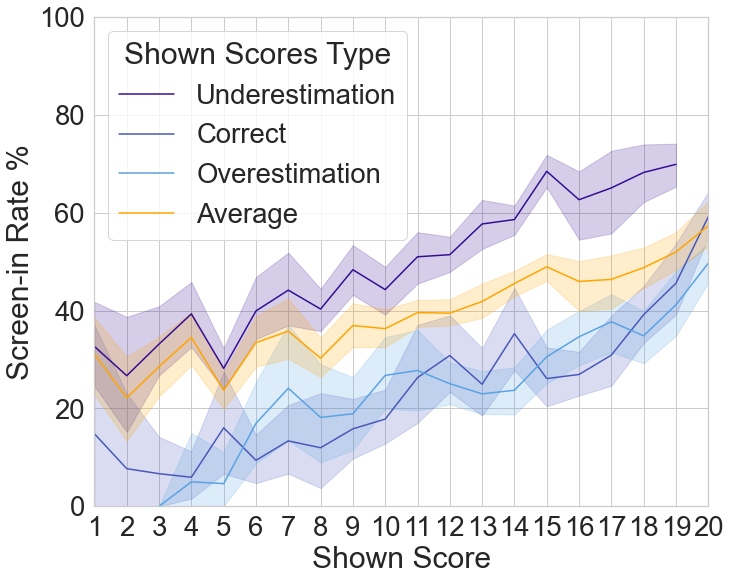}
    \caption{Screen-in Rate by Types of Shown Scores. Screen-in rate is higher when shown scores are underestimates of the assessed scores compared to when they are correct or overestimates.}
    \label{fig:screen_in_by_glitch}
  \end{minipage}
\end{figure}

Having summarized the case mixture encountered by the different screener types, we now turn to questions concerning how screener decisions varied with the AFST score.  Figure~\ref{fig:screen_in_by_screener} shows a plot of screen-in rates for different values of the shown score.  Interestingly, although screeners face similar distributions of AFST scores (both shown and assessed), senior screeners screen in at a higher rate than both junior and temp screeners at each value of the shown score. Assuming the predicted risk is a good indication of the real risk of child maltreatment, one should expect similar screen-in rates instead of consistent gaps among screener types. This suggests that screeners may have made different judgement calls on referrals of the same shown scores depending on their experiences.  

On the basis of this plot alone, we cannot rule out the possibility that difference in screen-in rate is due to dimensions of case risk or complexity that are not captured by the shown AFST score.  In our regression analyses we probe this question in more depth by explicitly controlling for other case characteristics available in the observed data.  Our results, which are reported in \textsection~\ref{sec:rq1_results}, are consistent with the findings described here.  


Figure~\ref{fig:screen_in_by_glitch} shows how screen-in rates varied across shown scores depending on whether the score was under-, over-, or correctly estimated.  If screeners made their decisions based solely on the shown scores, we would expect the screen-in rates to be similar regardless types of display error.  Instead we see that screen-in rates were significantly higher when the shown score underestimated the assessed score, but were comparable when the shown score was an over-estimate or agreed with the assessed score.  
This finding suggests that screeners use available external data sources to make their own assessments, and adjust for any perceived discrepancy in shown scores.

Both observations about screeners raise interesting questions about why they screen in at different rates and how they mitigate the effects of the glitch in real time.

\subsection{Race}
\label{sec:data_desc_race}
In this section we describe our findings surrounding how the AFST scores, screening decisions, and exposure to the glitch varied across racial groups.  According to 2019 population estimates provided by the US Census Bureau \cite{census:2019}, 13.4\% of Allegheny County residents are Black or African-American alone, and 2.3\% are mixed-race.  79.9\% of residents are White alone.  It is therefore not surprising that
Allegheny County CYF receives more referrals on white children than on any other racial group.  However, black children are disproportionately represented among children referred to CYF relative to their presence in the resident population. Using a black-prioritized race coding, we observe that 43.5\% of the referrals received post-AFST implementation concern white children, compared to 39.6\% concerning black children, 7.9\% concerning children of two or more races, 3.5\% concerning children of unknown race, and 5.2\% with children coming from different racial background (eg. a multi-racial household where one child is identified as a child of two or more race and another child is identified as white). DHS does not define other racial categories. We focus on referrals associated with exclusively white and black children because they represent a majority of the referrals.

\begin{figure}[h]
  \begin{minipage}[t]{0.45\textwidth}
   \centering
    \includegraphics[width=\textwidth]{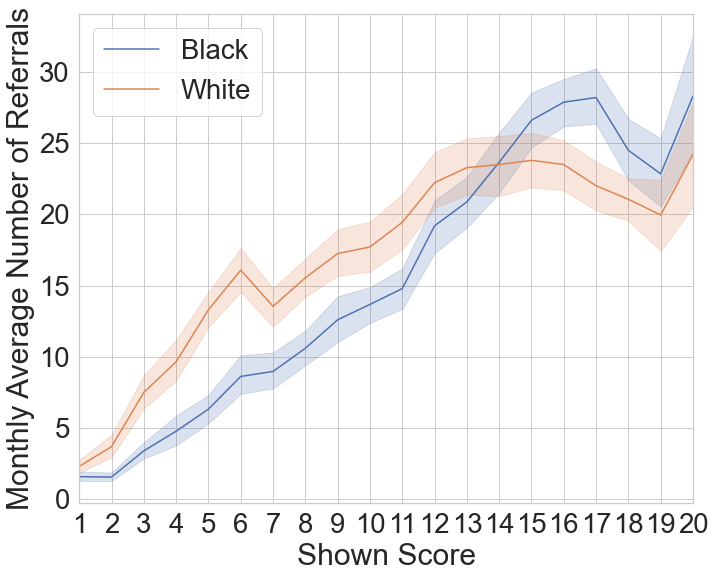}
    \caption{Monthly Number of Referrals by Race. We see that a majority of the higher risk referrals (score $> 14$) concern black children. Note that only around 13\% of residents in Allegheny County are black. Error bands correspond to 95\% pointwise confidence intervals.}
    \label{fig:num_referral_by_race}
  \end{minipage}
  \hfill
  \begin{minipage}[t]{0.45\textwidth}
    \includegraphics[width=\textwidth]{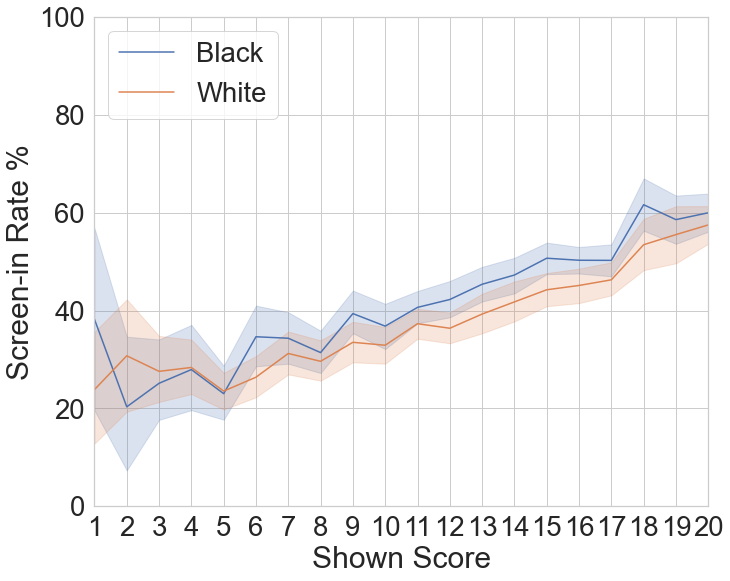}
    \caption{Monthly Screen-in Rate by Race. At almost every shown risk score, referrals on black children are screened in at a higher rate than those on white children.  Error bands correspond to 95\% pointwise confidence intervals.}
    \label{fig:screen_in_by_race}
  \end{minipage}
\end{figure}

Figure~\ref{fig:num_referral_by_race} shows the average number of referrals received each month for each racial group across the different score levels.  We find that there are more low-scoring referrals on white children than black children. There are slightly more referrals on black children than white children for risk scores above 14. 

Figure~\ref{fig:screen_in_by_race} shows the screen-in rate across shown AFST score for black and white children.  Assuming shown scores are a good reflection of case risk, one would expect the screen-in rate to increase as the risk score increases.  Furthermore, the scores are \textit{equally} good reflections of case risk across racial groups, one would also expect the screen-in rate at each score to be roughly the same across racial groups.  
We observe that for both black and white children, their chance of being screened in increases as the risk score increases.
However, at almost every risk score, referrals on black children are screened in at a higher rate than referrals on white children. This observation raises potential concerns about "over screening" of black children, which was raised previously by \citet{_rauktis_role_2010}.

\begin{figure}[h]
\centering
\includegraphics[scale=0.3]{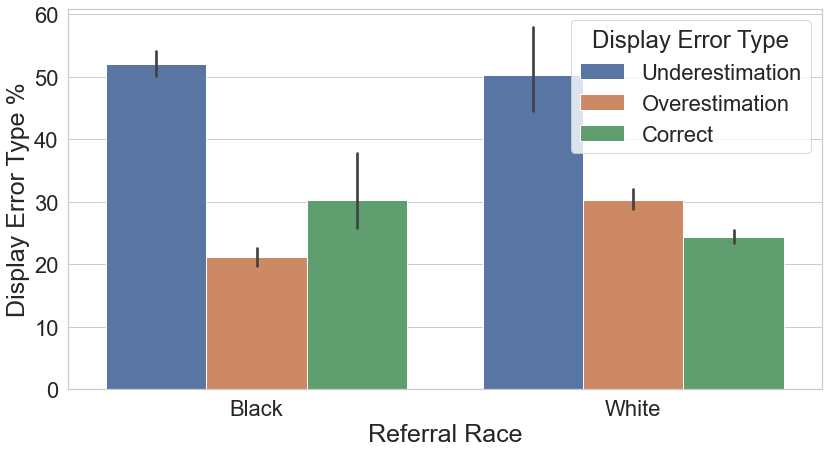}
\caption{Display Error Distribution By Race. Referrals associated with white children are more likely to have overestimated shown scores than the ones on black children. Error bars are 95\% confidence intervals.}
\label{fig:error_by_race}
\end{figure}

Figure~\ref{fig:error_by_race} shows a breakdown of the correspondence between the shown score and assessed score for the two racial groups.  
We observe that, consistent with the general direction of misestimation, both races are on average more likely to have shown scores that are underestimates of the assessed scores. However, referrals on white children are more likely than black children to have overestimated shown scores. Therefore, part of the explanation for why white referrals are screened in at a lower rate than black ones conditioned on the same risk could be that the shown scores for white referrals are in fact more likely to be overestimates of the assessed risk. We address this issue by controlling for types of display errors in RQ3. 

\subsection{Summary of Descriptive Analyses}
\label{sec:data_desc_summary}
These descriptive analyses show that there is clear variation in the association between screening decisions, screener experience, and race. Despite receiving referrals of similar distributions of AFST-assessed (and shown) risk, senior screeners are more likely to screen in referrals at almost any given shown risk than junior and temp screeners. Screeners overall screen in referrals at a much higher rate if the shown scores are underestimations of the assessed risk. Lastly, black referrals are consistently screened in at a higher rate than white referrals with the same shown scores. Race also appears to be related to the exposure to different types of display errors. 

\section{Quantitative Research Methodology}
\label{sec:research_method}
Together, these initial findings motivate us to investigate how screener experience could affect their use of algorithmic assessment and their ability to mitigate misestimation due to the glitch, and whether there is evidence of racial disparities in screen in rates, potentially driven by these mechanisms.  In this section we more formally state our three main research questions and describe the econometric quantitative methods that we use to address them. 

\subsection{Research Questions}
\label{sec:research_method_RQs}
The three main research questions are:
\begin{itemize}
\item RQ1 Effect of Work Experience on Algorithm Utilization: How does call screeners' experience affect their use of shown scores?  In particular, are the decisions of senior screeners more or less aligned than the decisions of junior screeners with the shown scores? Likewise, are the decisions of full-time screeners more or less aligned than the decisions of part-time screeners with the shown scores?
\item RQ2 Tendency to Override When Shown Scores Differ from Assessed Scores: 
    \begin{itemize}
    \item RQ2.1 ($\tilde{M} = 0$, cases require no mandatory screen-in) Does experience influence screeners' tendency to screen in a referral given incorrect displays of assessed scores? 
    \item RQ2.2 ($\tilde{M} = 1$, cases require mandatory screen-in and supervisor override) When supervisors make decisions to override mandatory screen-ins, do their tendencies to do so vary with the experience of the screeners who make the screening recommendations?
    \end{itemize}
\item RQ3 Racial Disparities in Screening Decisions by Work Experience Given the Same Predicted Risk: Are there racial disparities in the screener's decisions when presented with referrals with the same predicted risk? 
\end{itemize}

Our analysis is aimed at identifying patterns and differences in patterns in the association between screening decisions and scores.  Due to the absence of good ground-truth maltreatment-related outcomes, we are unable to analyse and do not speak to whether observations of greater or lesser alignment with the AFST score corresponds to better or worse decision-making.  Our study examines patterns in decision-making, not decision quality.  Examining decision quality is a very interesting question but is beyond the scope of the current study, and beyond what our available data enables us to effectively assess.  

\subsection{Empirical model}
\label{sec:research_method_models}

In this section we describe the econometric quantitative methods that underpin our primary analysis to answer the three main research questions.  Taking the screening decision as the outcome, we estimate linear probability models with robust clustered standard errors \cite{cameron2015practitioner} at the screener level. In our robust clustered standard errors model, we assume the model errors in different time periods for a given screener may be correlated (within-screener correlation across decisions), but model errors for different screeners are assumed to be uncorrelated. The linear models allow us to directly interpret the coefficients to address the main research questions. The notations are listed in Table ~\ref{table:notation}.

We now provide the model specifications that we use to answer our three main request questions.  To avoid a proliferation of notation, the expressions below are written in short-hand to highlight the control variables and interaction terms. We clarify this point further below. We use STATA for conducting the analyses, which automatically drops variables that are perfectly collinear with others in estimating coefficients. Full equations and explanations are provided in Appendix ~\ref{section:appendix_full_methodology}. 

\subsubsection{RQ1 Effect of Work Experience on Algorithm Utilization} 
\label{sec:research_method_RQ1}

Senior and junior workers are full-time staff who regularly and consistently interact with AFST. Senior staff also have experience in making screening decisions without AFST, which trained them to access and interpret data on the referrals from a variety of data sources in ClientView. Although domain expertise can aid one's judgement, it can also be negatively associated with utilization of algorithmic assessment \cite{Burton20, Logg2019AlgorithmAP}. Thus, we might expect senior screeners to rely less on the AFST assessments than do junior and temp screeners. 

Using data post December 2016, and senior screeners as the reference category, we specify whether a referral $i$ is screened in under supervisor $j$ at time $t$ as a function of the shown scores, screener types, and the interaction between shown scores and screener types. Coefficients of interests are on the interaction term, which show the extent to which screeners use shown scores as they increase. We additionally control for referral characteristics, month and supervisor fixed effects:
{\begin{equation} \label{eq:1}  
    Y_{i,t,j} = \beta_0 + \beta_1 \tilde{S}_i + \beta_2 screenerType_{i,t} + \beta_3 screenerType * \tilde{S}_{i,t} + \theta_1 X_{i,t} + \gamma_t + \omega_j + \epsilon_{i,t,j}
\end{equation}}

Readers are reminded that a full notation table is provided in Table~\ref{table:notation}.   As mentioned earlier, we are taking some notational shortcuts here.  For instance, $screenerType_{i,t}$ can take on values senior, junior, and temp.  So the expression $\beta_2 screenerType_{i,t}$ technically treats $\beta_2$ as a vector of coefficients, and $screenerType_{i,t}$ as a vector of indicator variables for each screener type.   Importantly, we remind readers that $X$ is a vector of variables on the characteristics of a referral including count, age, and gender of the victims and perpetrators, total and unique counts of allegations, specific types of alleged abuses (out of the 237 categories), and whether the referral was made by a mandatory reporter.  These are control variables that we understand to be many of the key factors screeners take into account when making screening decisions.  We illustrate differences in types of the most common reported alleged abuse and their screen-in rates between black and white referrals in Appendix ~\ref{section:appendix_summary_graphs_referral_char}.  

\subsubsection{RQ2 Tendency to Override When Shown Scores Differ from Assessed Scores}
\label{sec:research_method_RQ2}

Previous work \cite{10.1145/3313831.3376638} has shown that screeners do not follow the shown scores blindly. Instead, they use other sources of information and are more likely to override algorithmic recommendations if the displayed risk is significantly miscalculated. 

As a follow up, we are interested in whether screeners' experience plays a role in their effectiveness in mitigating the display glitch, and if supervisors' decisions to override mandatory screen-in depend on the experience of the screeners who make the screening recommendation.

For this analysis, we use referrals between December 1$^{st}$, 2016 and July 31$^{st}$, 2018 where both shown and assessed scores are available. Figure~\ref{fig:num_referral_by_race} has suggested that when the shown score is an underestimation of the assessed score, the screen-in rate will be higher than when the shown score is correctly displayed or overestimated. Thus we define $glitchType$ as a categorical variable denoting if the shown score is an under- or over- or correct estimation of the assessed score. 

First, we check if glitch type is associated with screening decisions:
{\begin{equation} \label{eq:2}
Y_{i,t,j} = \beta_0 + \beta_1 \tilde{S}_{i,t} + \beta_2 glitchType_{i,t} + \theta_1 X_{i,t} + \gamma_t + \omega_j + \epsilon_{i,t,j}
\end{equation}}

We then study if screeners respond differently to each type of glitch: 
{\begin{equation}\label{eq:3}
Y_{i,t,j} = \beta_0 + \beta_1 \tilde{S}_{i,t} + \beta_2 screenerType_{i,t} + \beta_3 glitchType * screenerType_{i,t} + \theta_1 X_{i,t} + \gamma_t + \omega_j + \epsilon_{i,t,j}.
\end{equation}}

The interpretation of the model above depends on whether supervisor deliberation is required. We run the same specification for when $\tilde{M}=0$ and when $\tilde{M}=1$ to answer RQ2.1 and RQ2.2 respectively.

\subsubsection{RQ3 Response to Race by Work Experience Given the Same Predicted Risk}
\label{sec:research_method_RQ3}

We focus on black and white referrals, since they constitutes the majority of the referrals. We subsequently drop referrals on other racial categories. Previous descriptive analysis shows that race seems to be associated with the exposure to the glitch. To study whether there are racial differences in how screeners respond to risk scores during the time period with the glitch, we use the following specification:
\begin{equation}\label{eq:4}
Y_{i,t,j} = \beta_0 + \beta_1 \tilde{S}_{i,t
} + \beta_2 Black_{i,t} + \beta_3 glitchType_{i,t} + \theta_1 X_{i,t} + \gamma_t + \omega_j + \epsilon_{i,t,j}.
\end{equation}
We then disaggregate the effect of race by screener types:
\begin{equation}\label{eq:5}
Y_{i,t,j} = \beta_0 + \beta_1 \tilde{S}_{i,t
} + \beta_2 screenerType_{i,t} + \beta_3 glitchType_{i,t} + \beta_4 Black * screenerType_{i,t} + \theta_1 X_{i,t} + \gamma_t + \omega_j + \epsilon_{i,t,j}.
\end{equation}
Lastly after the glitch was fixed, we check if a particular type of screener is more receptive to the influence of race:
\begin{equation}\label{eq:6}
Y_{i,t,j} = \beta_0 + \beta_1 \tilde{S}_{i,t
} + \beta_2 screenerType_{i,t} + \beta_3 Black * screenerType_{i,t} + \theta_1 X_{i,t} + \gamma_t + \omega_j + \epsilon_{i,t,j}.
\end{equation}

\subsubsection{Discussions of the linear probability model.}
Even though the outcome variable, the screening decision, is binary, our primary analysis is conducted using a Linear Probability Model (LPM) to provide ease of interpretation.  There are several concerns that arise when using an LPM to model binary data: (i) the model may be a poor fit when the relationship between continuous inputs and the outcome is clearly sigmoidal and non-linear; (ii) the standard LPM assumes heteroskedastic errors; and (iii) the model may produce probability estimates outside of the unit interval.   

Regarding (i) and (iii), we observe in Figures ~\ref{fig:screen_in_by_screener}, ~\ref{fig:screen_in_by_glitch}, and  ~\ref{fig:screen_in_by_race}, that screen-in rates appears to increase approximately linearly with the score, and that the rates tend to fall in the 20-80\% range.  This suggests that the linear form of the LPM is appropriate, and that probability estimates aren't expected to frequently fall out-of-bounds.  We would expect linear models and logistic regressions to  fit about equally well in this regime \cite{von_hippel_2019}. Thus, we favor linear probability models for its interpretability.

To address (ii), we rely on robust standard errors to ensure that we capture the correct variance structure when calculating p-values. 

Finally, as a robustness check, we reproduce the main results using logistic regression in Appendix ~\ref{section:appendix_logit}. We find that our qualitative results and findings of statistical significance are robust to model specification.

\section{Results}
\label{sec:results}

In this section we present the results of our quantitative analysis and interpret the findings in the context of our three primary research questions.  Our regression results are captured in tables such as Table~\ref{table:utilization_of_scores_by_screener}, the format of which may be unfamiliar to readers.  These tables present regression coefficients for key variables of interest (defined by the names in the first row of the table) across several model specifications (presented as the columns).  So, for instance, column (1) shows coefficient estimates and standard errors (in parentheses) for the shown score $\tilde S$, worker type, and score-by-worker-type interactions in a model specification that does not include Month Fixed Effects (FE), Supervisor FE, or Referral Characteristics ($X$).  Column (3) presents this same information with all fixed effects and referral characteristics, $X$, also included in the regression.  

\subsection{RQ1 Effect of Work Experience on Algorithm Utilization} \label{sec:rq1_results}
Table ~\ref{table:utilization_of_scores_by_screener} shows that after the AFST implementation, shown scores are highly associated with screen-in decisions. In column (3) and (4), where we control for referral characteristics, we observe that a one-point increase in shown score is statistically significantly associated with a 0.8\% increase in a referral's chance of being screened in, all else equal. We note that these results remain statistically significant when the data is analysed using logistic regression, as shown in Appendix ~\ref{section:appendix_logit}. Given that the score can range from 1 to 20, this means that even when two referrals share the same demographic and allegation characteristics, a referral that is shown to have a score of 20 is 15.2\% more likely (in absolute terms) to be screened than a referral with a shown score of 1. In reality, cases that are at opposite ends of the shown score scale are unlikely to share all other modelled characteristics in common.  It is more likely that two otherwise similar cases might differ in their shown AFST score by 2-5 points, which would correspond to a 1.6 - 4.0\% difference in screen-in rates.  Considering that the County receives nearly 15,000 referrals each year, these differences can amount to a large number of affected cases.  To put these figures into further perspective, we note that \citet{green2021algorithmic} in a laboratory setting observed that use of a criminal risk assessment tool reduced the likelihood of pretrial detention by an average of 2.4\% .  This is comparable, at least in order of magnitude, to the association identified in our analysis. 
\begin{table}[]
\caption{SCREENERS' UTILIZATION OF RISK SCORES}
\centering
\scalebox{0.85}
{
\def\sym#1{\ifmmode^{#1}\else\(^{#1}\)\fi}
\begin{tabular}{l*{4}{c}}
\hline\hline
            &\multicolumn{1}{c}{(1)}&\multicolumn{1}{c}{(2)}&\multicolumn{1}{c}{(3)}&\multicolumn{1}{c}{(4)}\\
\hline
Shown Score &      0.0184\sym{***}&      0.0131\sym{***}&      0.0078\sym{***}&      0.0078\sym{***}\\
            &    (0.0011)         &    (0.0012)         &    (0.0009)         &    (0.0009)         \\
[1em]
Junior &     -0.1655\sym{***}&     -0.0834\sym{**} &     -0.0995\sym{***}&     -0.0979\sym{***}\\
            &    (0.0271)         &    (0.0254)         &    (0.0215)         &    (0.0214)         \\
[1em]
Temp &       -0.1194\sym{***}&     -0.0458         &     -0.0318\sym{*}  &     -0.0308         \\
            &    (0.0320)         &    (0.0250)         &    (0.0157)         &    (0.0156)         \\
[1em]
Junior * Shown Score &      0.0035\sym{*}  &      0.0030         &      0.0031\sym{*}  &      0.0030\sym{*}  \\
            &    (0.0017)         &    (0.0015)         &    (0.0014)         &    (0.0014)         \\
[1em]
Temp * Shown Score &     0.0001         &      0.0003         &     -0.0007         &     -0.0008         \\
            &    (0.0016)         &    (0.0014)         &    (0.0013)         &    (0.0013)         \\
[1em]
Month FE &   &  Y   &  Y  & Y\\
[1em]
Supervisor FE &   &  Y   &  Y & Y \\
[1em]
Referral Characteristics  &   &     &  Y   &      \\
[1em]
Referral Characteristics (No Race) &   &     &  & Y\\
\hline
\(N\)         &       28,736         &       28,736         &       28,736         &       28,736         \\
\hline\hline
\multicolumn{5}{l}{\footnotesize Referrals of all races are included.}\\
\multicolumn{5}{l}{\footnotesize The dependent variable is a binary variable, 1 if screened in and 0 otherwise.}\\
\multicolumn{5}{l}{\footnotesize The mean of the dependent variable is 0.4244, meaning the average screen in rate is 42.44.}\\
\multicolumn{5}{l}{\footnotesize Robust standard errors clustered by screener in parentheses. }\\
\multicolumn{5}{l}{\footnotesize \sym{*} \(p<0.05\), \sym{**} \(p<0.01\), \sym{***} \(p<0.001\)}\\
\end{tabular}
}
\label{table:utilization_of_scores_by_screener}
\end{table}

When we compare specifications (2) and (3), we observe that the effect of shown scores decreases by 0.5\% but still remains significant, which reflects the fact that that shown scores contain information that overlaps with but is also distinct from the modelled referral characteristics, and screeners' decisions are significantly aligned with the shown score along dimensions not explained by the modelled referral characteristics alone.

Interestingly, even though junior screeners have considerably lower screen-in rates than senior screeners, we find that their decisions are significantly more strongly associated to the risk scores.  Whereas each 1-point difference in risk score, all else equal, corresponds on average to a 0.8\% greater likelihood of screen-in by a senior worker, this increases to $0.8 + 0.31 = 1.11\%$ for junior workers.  
Although the practical difference appears small, the estimates are stable across all specifications. This is consistent with a hypothesis that junior screeners rely slightly more heavily on AFST when making screening decisions compared to senior screeners, who had experience performing the same decision-making task without AFST.  

To compare full-time and part-time staff, we test the equality of coefficients between junior and temp screeners using our preferred specification under column (3). We observe statistically significant difference ($F(1,75)=6.06, Prob>F=0.0161$) between them at the 0.05 significance level. However, there is no statistically significant difference between senior and temp screeners. This observation contradicts our hypothesis that senior and temp screeners' decisions should be differently aligned with the shown scores due of their more frequent interactions with AFST. At the same time, we may also anticipate a non-linear effect of experience on screening decision if senior screeners increasingly weigh their subjective judgement more than AFST assessments. Thus, the similarity between senior and temp screeners can potentially be explained by a hypothesis that either the heavier weights on subjective judgement or infrequent encounters of AFST in decision-making could reduce usage of algorithmic risk assessment. However, we are unable to test this hypothesis directly in the present study and conclude we do not observe significant difference in usage of AFST between senior and temp screeners.

In summary, although AFST uses some information that is routinely available to and is viewed by call screeners at time of screening, it also synthesizes from other data sources in a manner that is significantly associated with screener decisions. Despite the fact that senior screeners screen in at a higher rate, junior screeners are more likely to screen in a case than senior and temp screeners, given the same referral characteristics and score. 

\subsection{RQ2 Tendency to Override When Shown Scores Differ from Assessed Scores} \label{sec:rq2_results}

Since screeners make different screening decisions given the same shown scores based on their experience, the question that then follows is, what happens if the shown scores are inaccurate displays of the assessed scores? 

The regression result in Table ~\ref{table:effect_of_glitch_on_screen_in} is consistent with the finding in Figure ~\ref{fig:screen_in_by_glitch}, which shows that when the shown scores are an underestimate of the assessed scores, call screeners are significantly more likely to screen in these referrals than when the scores are correctly displayed. The effect of overestimation appears similar to if not slightly below that of correct estimation. This suggests that screeners may consider perceived discrepancy between self-assessment of referral risk and shown score risk in the screening process.

\begin{table}[]
\caption{ESTIMATED EFFECTS OF DISPLAY DISCREPANCY ON SCREEN-IN}\centering\scalebox{0.85}
{
\def\sym#1{\ifmmode^{#1}\else\(^{#1}\)\fi}
\begin{tabular}{l*{4}{c}}
\hline\hline
            &\multicolumn{1}{c}{(1)}&\multicolumn{1}{c}{(2)}&\multicolumn{1}{c}{(3)}&\multicolumn{1}{c}{(4)}\\
\hline
Shown Score &    0.0263\sym{***}&      0.0163\sym{***}&      0.0114\sym{***}&      0.0116\sym{***}\\
            &    (0.0009)         &    (0.0008)         &    (0.0009)         &    (0.0009)         \\
[1em]
Overestimated &   -0.0193         &     -0.0178\sym{*}  &     -0.0366\sym{***}&     -0.0361\sym{***}\\
            &    (0.0099)         &    (0.0085)         &    (0.0085)         &    (0.0084)         \\
[1em]
Underestimated &      0.2721\sym{***}&      0.1580\sym{***}&      0.1225\sym{***}&      0.1229\sym{***}\\
            &    (0.0113)         &    (0.0107)         &    (0.0099)         &    (0.0100)         \\
[1em]
Month FE &   &  Y   &  Y  & Y\\
[1em]
Supervisor FE &   &  Y   &  Y & Y \\
[1em]
Referral Characteristics  &   &     &  Y   &      \\
[1em]
Referral Characteristics (No Race) &   &     &  & Y\\
\hline
\(N\)       &       14,475         &       14,475        &      14,475         &       14,475        \\
\hline\hline
\multicolumn{5}{l}{\footnotesize Referrals of all races between 12/1/16-31/7/18 where both shown and assessed scores are included.}\\
\multicolumn{5}{l}{\footnotesize The dependent variable is a binary variable, 1 if screened in and 0 otherwise.}\\
\multicolumn{5}{l}{\footnotesize The mean of the dependent variable is 0.4295, meaning the average screen in rate is 42.95}\\
\multicolumn{5}{l}{\footnotesize Robust standard errors clustered by screener in parentheses. }\\
\multicolumn{5}{l}{\footnotesize \sym{*} \(p<0.05\), \sym{**} \(p<0.01\), \sym{***} \(p<0.001\)}\\
\end{tabular}
}
\label{table:effect_of_glitch_on_screen_in}
\end{table}

\subsubsection{RQ2.1 Non-mandatory Screen-in} \label{sec:rq2_1_results}

Table ~\ref{table:non_madatory_screen_in_by_screener} shows that for referrals that are not mandated to be screened in, screeners' decisions are not significantly different when shown scores are overestimates compared to when they are correct. However, screen-in rates are higher when shown scores are underestimates of the assessed scores. There is no statistically significant difference in screeners' tendency to adjust screening decisions $(F(2,59)=0.10, Prob>F=0.9075)$ across types of experience when shown scores are underestimates (specification (3)).

\begin{table}[]
\caption{SCREEN-IN OF NON-MANDATORY REFERRALS BY SCREENER AND DISPLAY DISCREPANCY}
\centering
\scalebox{0.8}
{
\def\sym#1{\ifmmode^{#1}\else\(^{#1}\)\fi}
\begin{tabular}{l*{4}{c}}
\hline\hline
            &\multicolumn{1}{c}{(1)}&\multicolumn{1}{c}{(2)}&\multicolumn{1}{c}{(3)}&\multicolumn{1}{c}{(4)}\\
\hline
Shown Score &        0.0230\sym{***}&      0.0139\sym{***}&      0.0090\sym{***}&      0.0091\sym{***}\\
            &    (0.0009)         &    (0.0007)         &    (0.0009)         &    (0.0009)         \\
[1em]
Junior &    -0.1567\sym{***}&     -0.0637\sym{**} &     -0.0712\sym{***}&     -0.0716\sym{***}\\
            &    (0.0215)         &    (0.0195)         &    (0.0192)         &    (0.0198)         \\
[1em]
Temp &    -0.1219\sym{***}&     -0.0207         &     -0.0178         &     -0.0181         \\
            &    (0.0261)         &    (0.0195)         &    (0.0174)         &    (0.0177)         \\
[1em]
Senior * Overestimation &    -0.0052         &      0.0010         &     -0.0198         &     -0.0193         \\
            &    (0.0136)         &    (0.0137)         &    (0.0150)         &    (0.0150)         \\
[1em]
Junior * Overestimation &      0.0443         &      0.0353         &      0.0171         &      0.0177         \\
            &    (0.0339)         &    (0.0248)         &    (0.0216)         &    (0.0218)         \\
[1em]
Temp * Overestimation &       0.0142         &      0.0083         &     -0.0147         &     -0.0144         \\
            &    (0.0139)         &    (0.0114)         &    (0.0104)         &    (0.0103)         \\
[1em]
Senior * Underestimation &      0.2798\sym{***}&      0.1764\sym{***}&      0.1380\sym{***}&      0.1383\sym{***}\\
            &    (0.0135)         &    (0.0157)         &    (0.0145)         &    (0.0146)         \\
[1em]
Junior * Underestimation &     0.2818\sym{***}&      0.1855\sym{***}&      0.1484\sym{***}&      0.1479\sym{***}\\
            &    (0.0362)         &    (0.0258)         &    (0.0197)         &    (0.0199)         \\
[1em]
Temp * Underestimation &      0.3117\sym{***}&      0.1819\sym{***}&      0.1402\sym{***}&      0.1405\sym{***}\\
            &    (0.0205)         &    (0.0170)         &    (0.0159)         &    (0.0159)         \\
[1em]
Month FE &   &  Y   &  Y  & Y\\
[1em]
Supervisor FE &   &  Y   &  Y & Y \\
[1em]
Referral Characteristics  &   &     &  Y   &      \\
[1em]
Referral Characteristics (No Race) &   &     &  & Y\\
\hline
\(N\)       &       12,981         &       12,981        &       12,981         &       12,981         \\
\hline\hline
\multicolumn{5}{l}{\footnotesize Referrals of all races between 12/1/16-31/7/18 where both shown and assessed scores are available}\\
\multicolumn{5}{l}{\footnotesize and no mandatory screen-in required are included.}\\
\multicolumn{5}{l}{\footnotesize The dependent variable is a binary variable, 1 if screened in and 0 otherwise.}\\
\multicolumn{5}{l}{\footnotesize The mean of the dependent variable is 0.4064, meaning the average screen in rate is 40.64.}\\
\multicolumn{5}{l}{\footnotesize Robust standard errors clustered by screener in parentheses. }\\
\multicolumn{5}{l}{\footnotesize \sym{*} \(p<0.05\), \sym{**} \(p<0.01\), \sym{***} \(p<0.001\)}\\
\end{tabular}
}
\label{table:non_madatory_screen_in_by_screener}
\end{table}

\subsubsection{RQ2.2 Mandatory Screen-in} \label{sec:rq2_2_results}
Interestingly, for the mandatory cases, it appears that as long as a referral passes the mandatory screen-in threshold, the shown scores are no longer significantly associated with screening decisions (considering other referral characteristics and fixed effects).

Similar to the non-mandatory cases, Table ~\ref{table:madatory_screen_in_by_screener} (specification (3)) shows that supervisors will be more likely to approve a mandatory screen-in if the shown score is underestimated compared to when it is correct regardless of the experience of screeners who work on the referral $(F(2,52)=0.46, Prob>F=0.6333)$. When the shown score is overestimated, however, supervisors seem to approve the referral at a lower rate if the referrals is associated with senior screeners. One possible explanation is that senior screeners may be better at presenting other information to justify not to screen in a referral where the shown score is overestimated. Nevertheless, we find no statistically significant difference $(F(2,52)=1.45, Prob>F=0.2439)$ among screeners under overestimation. Thus, we find no conclusive evidence showing that there is sufficiently significant differences in how screeners experience would be associated with supervisors' decisions to approve or override mandatory screen-ins.

\begin{table}[]
\caption{SCREEN-IN OF MANDATORY REFERRALS BY SCREENER AND DISPLAY DISCREPANCY}
\centering
\scalebox{0.7}
{
\def\sym#1{\ifmmode^{#1}\else\(^{#1}\)\fi}
\begin{tabular}{l*{4}{c}}
\hline\hline
            &\multicolumn{1}{c}{(1)}&\multicolumn{1}{c}{(2)}&\multicolumn{1}{c}{(3)}&\multicolumn{1}{c}{(4)}\\
\hline
Shown Score &        0.1051\sym{***}&      0.0303\sym{*}  &      0.0229         &      0.0229         \\
            &    (0.0192)         &    (0.0142)         &    (0.0130)         &    (0.0133)         \\
[1em]
Junior &      -0.1620\sym{*}  &     -0.1092         &     -0.1089         &     -0.1092         \\
            &    (0.0680)         &    (0.0647)         &    (0.0550)         &    (0.0557)         \\
[1em]
Temp &      -0.1035         &     -0.0126         &     -0.0019         &     -0.0035         \\
            &    (0.0602)         &    (0.0256)         &    (0.0236)         &    (0.0239)         \\
[1em]
Senior * Overestimation &     -0.0805         &     -0.0632\sym{*}  &     -0.0523\sym{**} &     -0.0524\sym{**} \\
            &    (0.0485)         &    (0.0259)         &    (0.0173)         &    (0.0162)         \\
[1em]
Junior * Overestimation &    -0.0268         &      0.0235         &      0.0167         &      0.0135         \\
            &    (0.0843)         &    (0.0589)         &    (0.0447)         &    (0.0453)         \\
[1em]
Temp * Overestimation &    0.0966         &      0.0046         &      0.0098         &      0.0082         \\
            &    (0.0636)         &    (0.0434)         &    (0.0393)         &    (0.0391)         \\
[1em]
Senior * Underestimation &      0.2524\sym{***}&      0.0549\sym{*}  &      0.0570\sym{*}  &      0.0577\sym{*}  \\
            &    (0.0304)         &    (0.0238)         &    (0.0224)         &    (0.0227)         \\
[1em]
Junior * Underestimation &     0.1715\sym{**} &      0.0223         &      0.0189         &      0.0203         \\
            &    (0.0528)         &    (0.0412)         &    (0.0381)         &    (0.0384)         \\
[1em]
Temp * Underestimation &     0.2538\sym{***}&      0.0650         &      0.0626         &      0.0621         \\
            &    (0.0691)         &    (0.0344)         &    (0.0354)         &    (0.0355)         \\
[1em]
Month FE &   &  Y   &  Y  & Y\\
[1em]
Supervisor FE &   &  Y   &  Y & Y \\
[1em]
Referral Characteristics  &   &     &  Y   &      \\
[1em]
Referral Characteristics (No Race) &   &     &  & Y\\
\hline
\(N\)       &        1,494         &        1,494         &        1,494         &        1,494         \\
\hline\hline
\multicolumn{5}{l}{\footnotesize Referrals of all races between 12/1/16-31/7/18 where both shown and assessed scores are available}\\
\multicolumn{5}{l}{\footnotesize and mandatory screen-in required are included.}\\
\multicolumn{5}{l}{\footnotesize The dependent variable is a binary variable, 1 if screened in and 0 otherwise.}\\
\multicolumn{5}{l}{\footnotesize The mean of the dependent variable is 0.6299, meaning the average screen in rate is 62.99.}\\
\multicolumn{5}{l}{\footnotesize Robust standard errors clustered by screener in parentheses. }\\
\multicolumn{5}{l}{\footnotesize \sym{*} \(p<0.05\), \sym{**} \(p<0.01\), \sym{***} \(p<0.001\)}\\
\end{tabular}
}
\label{table:madatory_screen_in_by_screener}
\end{table}

\begin{figure}[t]
  \begin{minipage}[t]{0.3\textwidth}
  \centering
    \includegraphics[width=\textwidth]{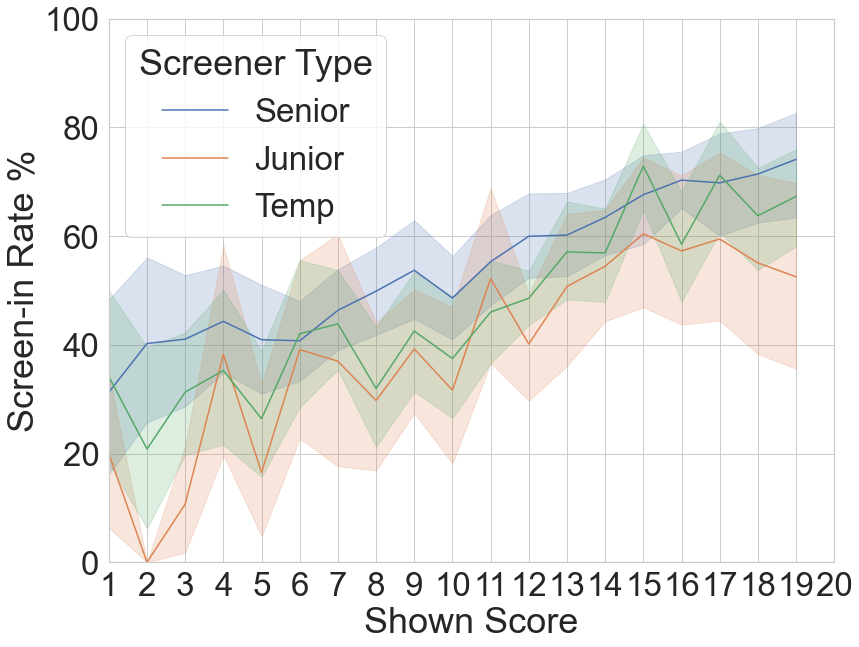}
     \subcaption{Underestimation}
  \end{minipage}
  \begin{minipage}[t]{0.3\textwidth}
  \centering
    \includegraphics[width=\textwidth]{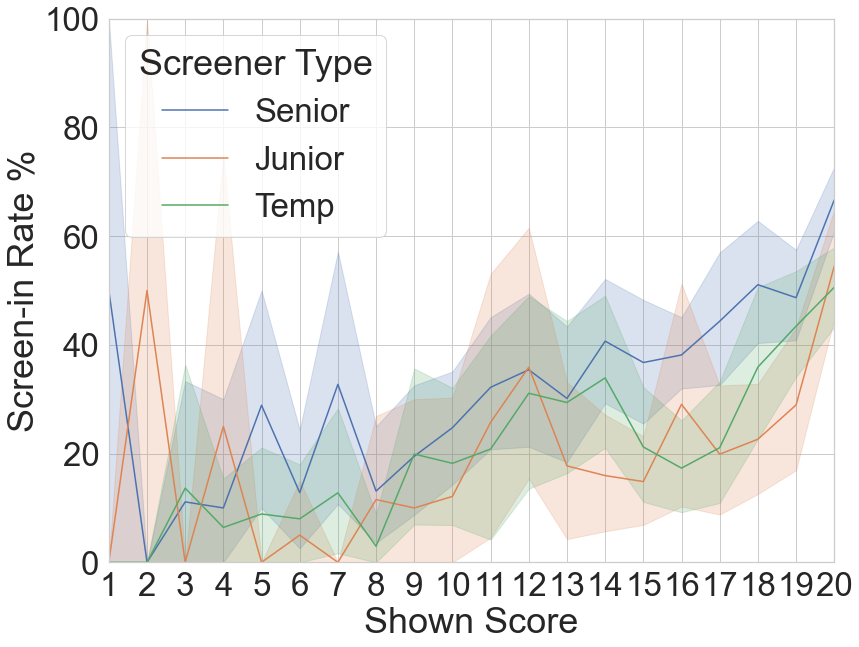}
    \subcaption{Correct Estimation}
  \end{minipage}
    \begin{minipage}[t]{0.3\textwidth}
  \centering
    \includegraphics[width=\textwidth]{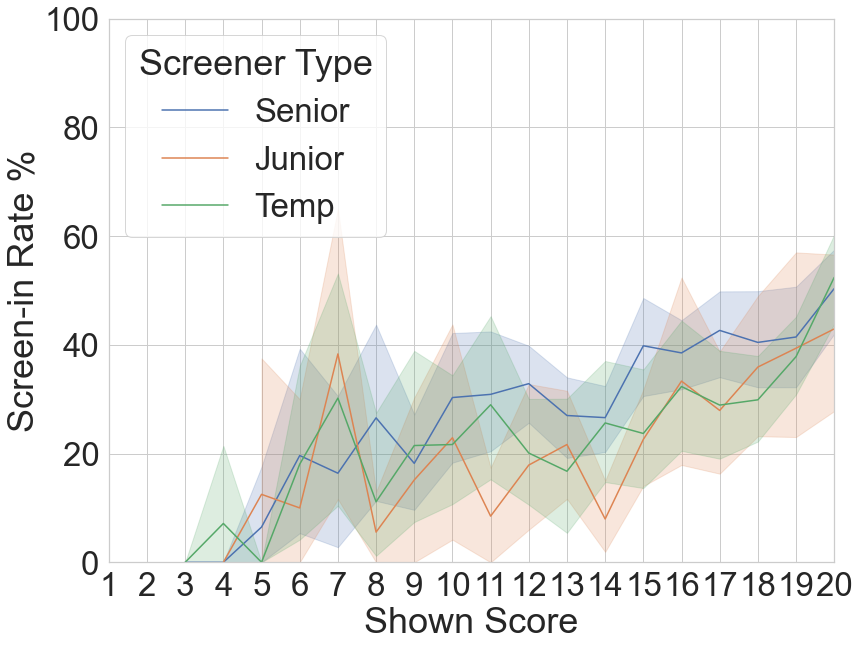}
    \subcaption{Overestimation}
  \end{minipage}
  \caption{Screen-in Rate of Non-Mandatory Referrals by Screener and Display Discrepancy. Conditioned on a type of display error, there is no differences among screeners' screen-in decisions based on experience.}
  \label{fig:screen_in_by_display_screener}
\end{figure}

In summary, we show that for both mandatory and non-mandatory referrals, when the shown score is an underestimation of the assessed score, the screen-in rate tends to be higher than when the shown score is correct. When the shown score is an overestimation of the assessed score, the screen-in rate tend to be similar or slight lower. Although one would reasonably expect more experienced screeners to better calibrate their screen-in decisions when the glitch occurs, overall, we find no conclusive evidence suggesting experience would significantly affect screeners or supervisors when adjusting screening decisions in presence of display errors as illustrated in Figure ~\ref{fig:screen_in_by_display_screener}. 

In addition to previous finding showing that screeners tend not to follow shown scores if they are misdisplayed, we show that effect of the glitch is more nuanced. Notably, the direction of display error is associated with different directions and sizes of the changes on screen-in rates. In particular, screeners decisions are consistent with what we would expect to see if they are capable of identifying when the shown scores are underestimates of the assessed scores by screening in these referrals at higher rates. This also suggests that staff overall may prefer reducing false negative errors rather than false positive errors by not screening out referrals simply because the displayed risk scores are low. On the contrary, it may be more difficult to identify overestimation errors, as the screen-in rates are rather similar between over and correct displays of assessed scores. Additionally, users were found to be more likely to change their prediction in the direction of the received machine advice in laboratory setting \cite{10.1145/3359280}. The difficulty in identifying estimation error and tendency to change prediction in alignment with algorithm advice together may contribute to the explanation of why overestimation errors can be "sticky." Understanding how these behaviors ultimately impact child welfare outcomes requires further investigation beyond the scope of the present analysis.

\subsection{RQ3 Response to Race by Work Experience Given the Same Predicted Risk} \label{sec:rq3_results}

Preliminary analyses in Section~\ref{sec:data_desc} show that, conditioned on shown scores, referrals associated with black children are screened in at a higher rate than white children. The subset of referrals where both shown and assessed scores are available further indicates that black and white children are disproportionately impacted by the display error. Thus, the observed racial disparity in the screen-in rate could be either due to screeners' use of race as a risk factor or the fact that the glitch impacts race differently. 

\begin{figure}[h]
  \begin{minipage}[t]{0.3\textwidth}
  \centering
    \includegraphics[width=\textwidth]{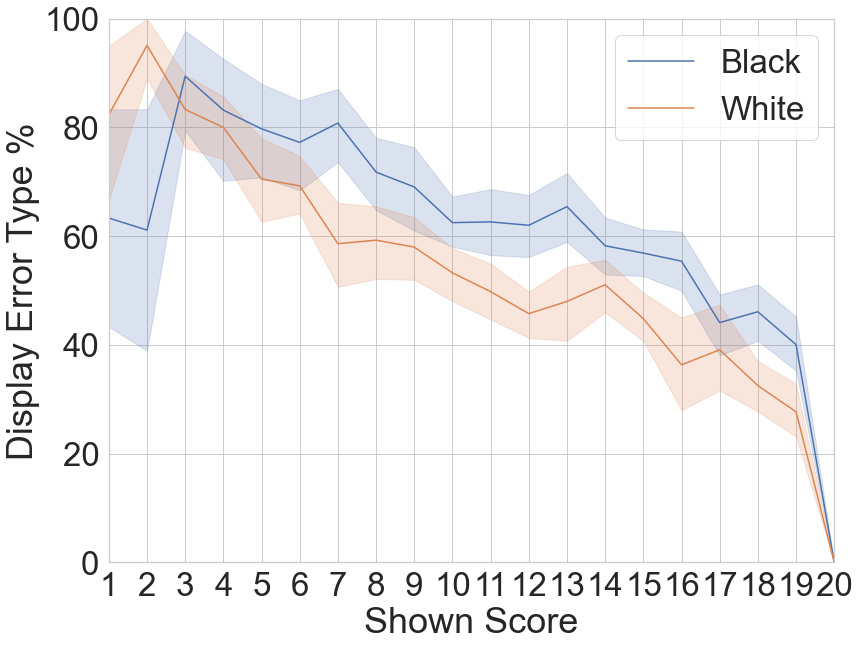}
     \subcaption{Underestimation}
  \end{minipage}
  \begin{minipage}[t]{0.3\textwidth}
  \centering
    \includegraphics[width=\textwidth]{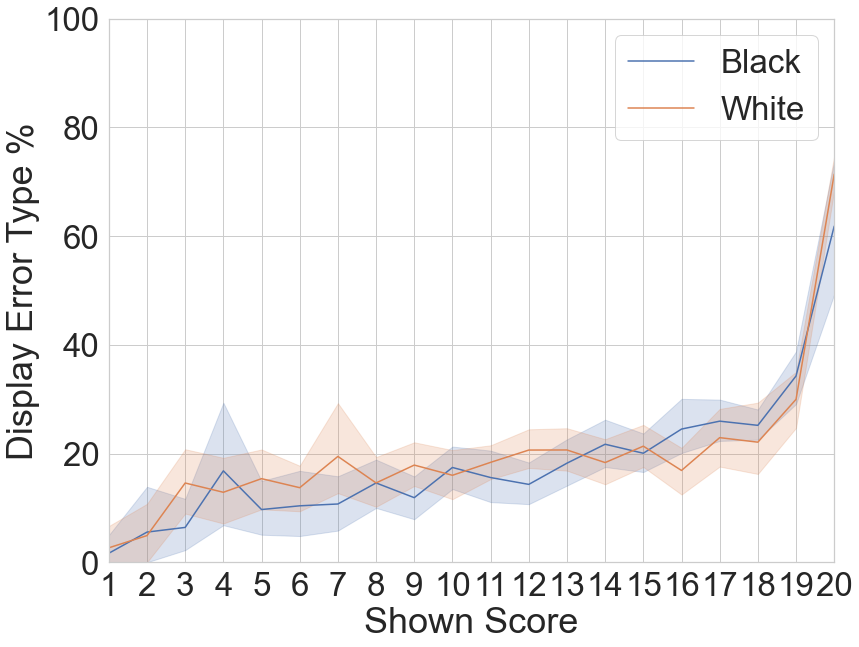}
    \subcaption{Correct Estimation}
  \end{minipage}
    \begin{minipage}[t]{0.3\textwidth}
  \centering
    \includegraphics[width=\textwidth]{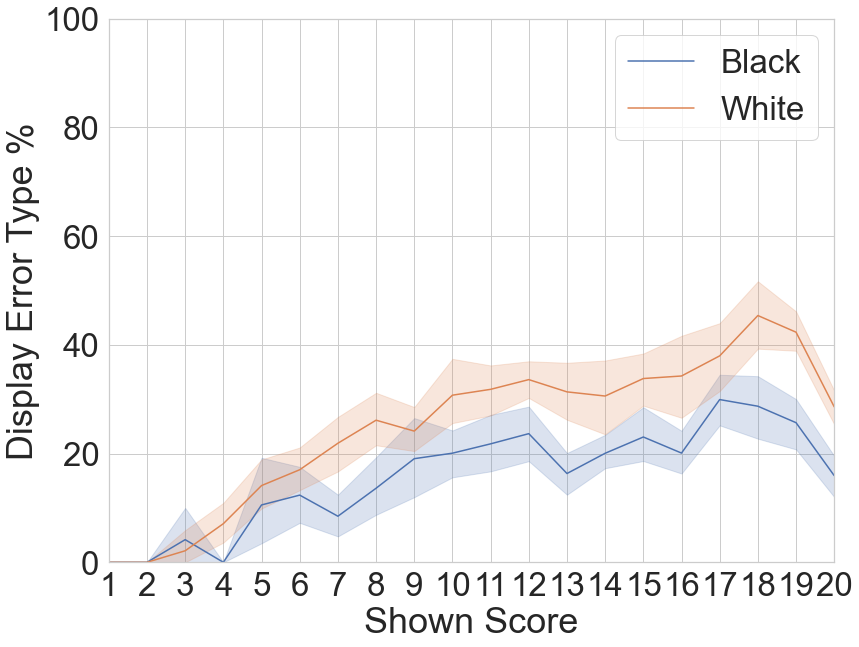}
    \subcaption{Overestimation}
  \end{minipage}
  \caption{Display Error Rate By Race. For a given shown score, the panel shows how likely the score is an over-/under-/correct estimation of the assessed score by race. The shown scores for referrals concerning white children were more likely than those for referrals concerning black children to be overestimates of assessed risk.}
  \label{fig:display_error_spectrum_race}
\end{figure}

Specifically, Figure ~\ref{fig:display_error_spectrum_race} shows that referrals on white children tend to show overestimated scores across the spectrum while referrals on black children tend to show underestimated scores.

\subsubsection{Prior To Glitch Fix} \label{sec:rq3_prior_glitch_result}
Distinct from prior literature, which evaluates this question without controlling for other potential explanatory variables, we show that race is not directly associated with screening decisions. The gap in screen-in rate between black and white children is attributable to the fact that the glitch affects the racial groups differently. Table ~\ref{table:race_discrepancy_on_screen_in} column (1) and (3) shows that during the glitch period, if a referral is associated with black children it is significantly more likely to be screened in conditioned only on shown scores or fixed effects. However, once we control for the effect of the glitch in (2) and (4), race is no longer significantly associated with the likelihood of being screened in.

Moreover, in column (5) and (6) where we control for the effects of other referral characteristics, race is not associated with screen-in decisions regardless whether the shown score was affected by the glitch. 
Because race is associated with being subject to a particular type of glitch as well as having differently distributed referral characteristics, we can explain the non-association between race and screen-in decision through either mechanism. 

\begin{table}[]
\caption{ESTIMATED EFFECTS OF RACE AND DISPLAY DISCREPANCY ON SCREEN-IN}
\centering
\scalebox{0.85}
{
\def\sym#1{\ifmmode^{#1}\else\(^{#1}\)\fi}
\begin{tabular}{l*{6}{c}}
\hline\hline
            &\multicolumn{1}{c}{(1)}&\multicolumn{1}{c}{(2)}&\multicolumn{1}{c}{(3)}&\multicolumn{1}{c}{(4)}&\multicolumn{1}{c}{(5)}&\multicolumn{1}{c}{(6)}\\
\hline
Shown Score &      0.0156\sym{***}&      0.0263\sym{***}&      0.0099\sym{***}&      0.0164\sym{***}&      0.0057\sym{***}&      0.0122\sym{***}\\
            &    (0.0010)         &    (0.0009)         &    (0.0009)         &    (0.0009)         &    (0.0009)         &    (0.0011)         \\
[1em]
Black       &      0.0353\sym{***}&      0.0021         &      0.0234\sym{***}&      0.0040         &      0.0120         &     -0.0023         \\
            &    (0.0089)         &    (0.0082)         &    (0.0061)         &    (0.0060)         &    (0.0066)         &    (0.0066)         \\
[1em]
Overestimation &        &     -0.0234\sym{*}  &                     &     -0.0196\sym{*}  &                     &     -0.0371\sym{***}\\
            &                     &    (0.0100)         &                     &    (0.0087)         &                     &    (0.0083)         \\
[1em]
Underestimation &                    &      0.2718\sym{***}&                     &      0.1567\sym{***}&                     &      0.1255\sym{***}\\
            &                     &    (0.0118)         &                     &    (0.0107)         &                     &    (0.0102)         \\
[1em]
Month FE &  & &  Y   &  Y  & Y & Y\\
[1em]
Supervisor FE &  & &  Y   &  Y & Y & Y \\
[1em]
Referral Characteristics (No Race)  &  & &  &     &  Y   &  Y    \\
\hline
\(N\)       &       12,019         &       12,019        &       12,019         &      12,019         &       12,019       &      12,019         \\
\hline\hline
\multicolumn{7}{l}{\footnotesize Referrals on black and white children between 12/1/16-31/7/18 where both shown and assessed scores are available are included}\\
\multicolumn{7}{l}{\footnotesize The dependent variable is a binary variable, 1 if screened in and 0 otherwise.}\\
\multicolumn{7}{l}{\footnotesize The mean of the dependent variable is 0.4350, meaning the average screen in rate is 43.50.}\\
\multicolumn{7}{l}{\footnotesize Robust standard errors clustered by screener in parentheses. }\\
\multicolumn{7}{l}{\footnotesize \sym{*} \(p<0.05\), \sym{**} \(p<0.01\), \sym{***} \(p<0.001\)}\\
\end{tabular}
}
\label{table:race_discrepancy_on_screen_in}
\end{table}

Disaggregating the result by screener experience, we find that the result is not driven by any particular type of screeners. Race is not associated with screeners' decisions. There is no statistically significant difference $(F(2,59)=1.09, Prob>F=0.3413)$ among screeners either using specification in column (6) as shown in Appendix~\ref{section:appendix_tables} Table~\ref{table:race_discrepancy_by_screener_glitch}. 

\subsubsection{Post-Glitch Fix} \label{sec:rq3_post_glitch_result}
As shown in Figure \ref{fig:implementation_changes}, the glitch was resolved on Feb 1$^{st}$ 2019, after which shown scores correctly reflected the assessed scores. Without the complication and using data post-glitch fix that includes referrals on black and white children between Feb 1$^{st}$ and Dec 31$^{st}$ 2019 (see Appendix ~\ref{section:appendix_summary_graphs_post_fix} for more details), again we find that race does not directly impact screening decisions, but is likely to do so indirectly through association with referral characteristics that are unevenly distributed by race. Table ~\ref{table:race_discrepancy_on_screen_in_post_glitch} shows that although race first appears to be associated with screening decisions conditioned on shown scores and fixed effects alone, this difference goes away once we control for referral characteristics (specification (3)).


\begin{table}[]
\caption{ESTIMATED EFFECTS OF RACE ON SCREEN-IN POST GLITCH}
\centering
\scalebox{0.85}
{
\def\sym#1{\ifmmode^{#1}\else\(^{#1}\)\fi}
\begin{tabular}{lccc}
\hline\hline
            &\multicolumn{1}{c}{(1)}&\multicolumn{1}{c}{(2)}&\multicolumn{1}{c}{(3)}\\
\hline
Shown Score &        0.0283\sym{***}&      0.0237\sym{***}&      0.0189\sym{***}\\
            &    (0.0013)         &    (0.0011)         &    (0.0013)         \\
[1em]
Black       &    0.0554\sym{***}&      0.0409\sym{***}&      0.0067         \\
            &    (0.0120)         &    (0.0104)         &    (0.0090)         \\
[1em]
Month FE & &  Y & Y \\
[1em]
Supervisor FE & &  Y & Y  \\
[1em]
Referral Characteristics (No Race)  & &   & Y \\
\hline
\(N\)       &        7,351         &        7,351         &       7,351        \\
\hline\hline
\multicolumn{4}{l}{\footnotesize Referrals on black and white children between 2/1/19-12/31/19 are included}\\
\multicolumn{4}{l}{\footnotesize The dependent variable is a binary variable, 1 if screened in and 0 otherwise.}\\
\multicolumn{4}{l}{\footnotesize The mean of the dependent variable is 0.4391 , meaning the average screen in rate is 43.91.}\\
\multicolumn{4}{l}{\footnotesize Robust standard errors clustered by screener in parentheses. }\\
\multicolumn{4}{l}{\footnotesize \sym{*} \(p<0.05\), \sym{**} \(p<0.01\), \sym{***} \(p<0.001\)}\\
\end{tabular}
}
\label{table:race_discrepancy_on_screen_in_post_glitch}
\end{table}

After explicitly controlling for possible explanatory variables, we find no statistically significant association between race and screening decisions. There is no difference $(F(2,49)=1.71, Prob>F=0.1908)$ among screeners in their consideration of race either, as illustrated in Appendix ~\ref{section:appendix_tables} Table ~\ref{table:race_discrepancy_by_screener_post_glitch}.

\section{Limitations} \label{sec:limitations}
To estimate the effects of AFST risk scores, we attempt to set up the empirical models as close to the decision contexts as possible. Nevertheless, we cannot fully capture the complexity of screeners' decision making process and make some broad assumptions in interpreting the results. 

Firstly, we assume the set of control variables represent factors workers would reasonably and effectively consider when making decisions in real time. The inclusion of those factors were based on our discussions with the call screeners and supervisors at Allegheny County CYF services. As discussed in \textsection~\ref{sec:background}, AFST synthesizes the county's integrated database that contains demographics, justice system involvement, public welfare histories, and behavioral health histories of child victims, parents, and alleged perpetrators. This represents a wider set of variables that were in our controls. Indeed, as we observe in \textsection~\ref{sec:rq1_results}
Table ~\ref{table:utilization_of_scores_by_screener}, when we control for case characteristics the coefficient on AFST decreases. This is expected as AFST contains information that partly overlaps with but is also unique from the controls. In reality, it is possible that call screeners may go above and beyond to look up other information in county's integrated database when making particular decisions. Our estimates will be biased to the extent to which this happens.

Secondly, we assume that "unobservables" -- factors that are available to the call screeners in real time, but are not available to us -- do not affect case assignment to workers. These may include factors such as the tone of the caller, verbal information passed on to workers, any background noises, and so on. We have shown in \textsection~\ref{sec:data_desc_case_mix} and Appendix ~\ref{section:appendix_summary_graphs_risk_distribution} and ~\ref{section:appendix_summary_graphs_post_fix}, AFST risk scores do not affect case assignments to workers. However, if the case assignment is driven by the unobservables, we may encounter a situation where senior workers are consistently assigned more risky cases, which provides an alternative explanation to why they screen in at a much higher rate than others.


Lastly, even though we find that race is not directly associated with screen-in decisions after we account for the differential exposure to the glitch and uneven distributions of referral characteristics, we acknowledge the historical impact of race and its effects on decisions through other channels. As a stylized example, if we assume that AFST fully replicates race \emph{and} if the true risk model is equal to race,  then we could get a model where the effect of race is severely underestimated if we estimate risk as a model of AFST and race. However, this would require assuming the true risk is fully represented by race which is unlikely the reality. Nonetheless, we may underestimate the effect of race to the extent that AFST which is based on historical records that can replicate racial bias, and are not claiming race cannot affect decisions in child welfare systems in any other ways.

\section{Discussion} \label{sec:discussion}
Using data on screening decisions following the deployment of the AFST, we investigated whether there is heterogeneity in the alignment between call screener decisions and the shown AFST score, and whether that heterogeneity may be resulting in racial disparities in screening decisions.  

We find that senior screeners persistently screen in referrals at higher rates than junior and temp screeners, given referrals with the same shown score and observed characteristics.  It is possible that this finding is driven by systematic differences in how supervisors assign cases to screeners.  We cannot rule out the possibility that supervisors are relying on information not reflected in our data to identify referrals that are more likely to warrant an investigation and preferentially assign those to more senior workers.  While we have controlled for many potential explanatory factors, and find no evidence of differences in AFST risk across the case mixes handled by different worker groups, it is nevertheless possible that factors known to supervisors but unavailable in our data explain the observed results.  

When the shown score is an incorrect display of the score that should have been assessed by the algorithm, screeners overall are more likely to screen in a referral if it is an underestimation of the assessed score, suggesting they are using information other than the score to compensate for the discrepancy between the shown scores and their own assessment of risk. However, the extent to which they do so is not associated with experience. For mandatory screen-in cases, supervisors similarly approve referrals with underestimated scores at higher rates regardless of the experience of the screeners who work on the referrals. These observations suggest that staff may be able to detect underestimation errors more easily than overestimation or prefer reducing false negative rather than false positive errors.

We find that race is not significantly associated with screening decisions. The gap in screen-in rates between black and white referrals is attributable to their differential exposure to the glitch and uneven distributions of referral characteristics. Likewise, we also find no racial differences in screen-in rates across worker experience levels once other observed explanatory factors are taken into account.  This finding does not indicate that there are no differences in screen-in rates across racial groups. It is clear that, in the aggregate, there are.  Rather, those differences are found to be attributable to other referral characteristics.  

We observe that the decisions of less experienced workers are more closely aligned with algorithmic risk scores than those of senior workers who had decision-making experience prior to the tool being introduced.  
System designers and agencies considering adopting algorithmic tools should therefore consider approaches to training and professional development that preserve institutional knowledge, particularly in high worker turnover domains such as child welfare call screening.

Policymakers can provide support to workers in understanding the goals, capabilities, and limitations of the risk assessment tool in relations to their own experiences, in order to achieve better reliance on the risk assessment and complementary between human and AI. For example, while the algorithm is developed for assessing the long-term risks of placement and re-referral, workers may focus more on imminent risks posed to children. Experienced workers may also pick up on cues that are not considered by the risk assessment tool, or interpret decision factors differently based on complex and changing social contexts. Such misalignment on goals and interpretation can lead to potential deflated trust and morale \cite{Bosk18}, and improper reliance on automation \cite{10.1145/3313831.3376813, Lee04, DZINDOLET2003697}. Thus, it is important that policymakers acknowledge both the unique strengths and weaknesses of workers and the risk assessment system, and develop strategies and training procedures that leverage and augment the strengths for better human-AI collaboration \cite{wilson2018collaborative, 9425540}. 

To address the constantly evolving needs and concerns of the child welfare systems, policymakers can foster a culture where workers feel valued and encouraged to provide candid feedback, and facilitate knowledge sharing within the organization to inform future iterations of the risk assessment system.




Future studies can further explore the relationship between screener, referral, and system characteristics to enrich our understanding of the changes in work dynamics in the presence of algorithms. An immediate follow-up would be to refine the notion of worker experience by decomposing experience into number of years of experience with the AFST and without, and then testing the hypothesis in algorithm aversion wherein experience with algorithmic decision aids is positively associated with the utilization of algorithmic judgement, while domain expertise is negatively associated with utilization \cite{Burton20, Logg2019AlgorithmAP}. Another direction points to promoting better matching between screeners and referrals. Past research on representative bureaucracy in child welfare has shown that when non-Caucasian caseworkers share the same racial/ethnic background as caregivers, caseworkers use more active strategies to connect caregivers to needed housing services \cite{doi:10.1086/675373}. Lastly, one may attempt to explore the implementation glitch as a natural experiment to answer questions about how the algorithm's development procedures could affect workers' trust and reliance on the system, though this is a challenging direction due to the non-random nature of the glitch \cite{10.1145/3313831.3376813, Lee04, DZINDOLET2003697}. Grounding these studies in institutional realities can produce useful insights for designing better human-AI collaboration.


\bibliographystyle{ACM-Reference-Format}
\bibliography{literature}

\newpage
\appendix
\section{Summary Graphs}
\label{section:appendix_summary_graphs}

\subsection{Distribution of Risk Scores}
\label{section:appendix_summary_graphs_risk_distribution}
The distributions of predicted and shown risks are similar among screeners for the subset where shown and assessed scores are available. 
\begin{figure}[h]
  \begin{minipage}[t]{0.48\textwidth}
   \centering
    \includegraphics[scale=0.25]{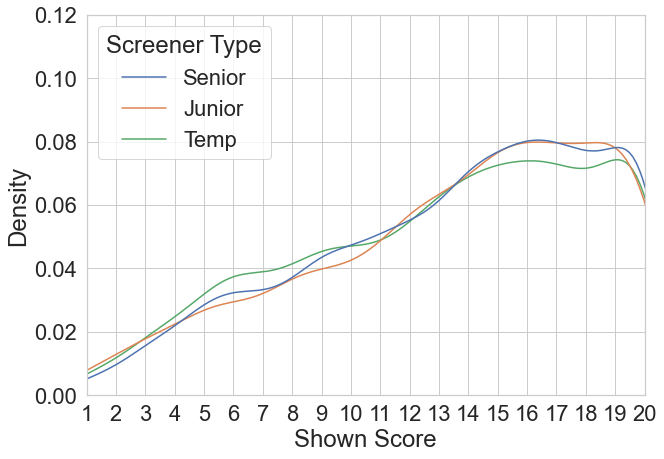}
    \\
    \subcaption{Shown Scores Distribution By Screener Type}
    \label{fig:subset_risk_1}
  \end{minipage}
  \hfill
  \begin{minipage}[t]{0.50\textwidth}
    \includegraphics[scale=0.25]{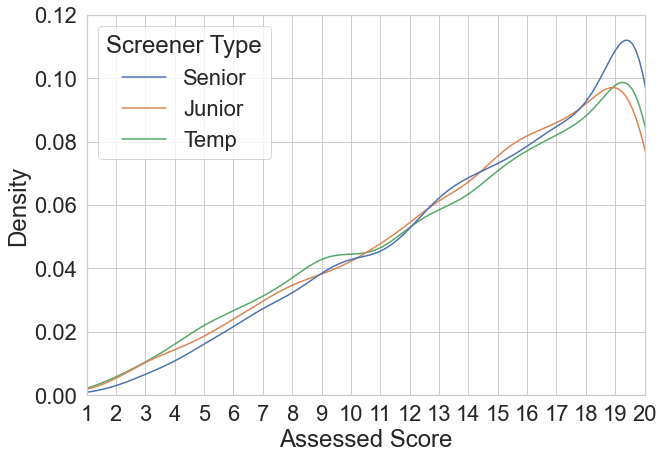}
    \\
     \subcaption{Assessed Scores Distribution By Screener Type}
      \label{fig:subset_risk_2}
  \end{minipage}
 \caption{Area under each screener type is normalized to be 1. Shown and assessed scores are similarly distributed among screener types. Shown and assessed scores distributions have different shapes but the differences are consistent across screener types.}
\end{figure}

\subsection{Post-Glitch Fix Dataset}
\label{section:appendix_summary_graphs_post_fix}
The post-glitch fix dataset contains referrals on black and white children from Feb 1$^{st}$ to Dec 31$^{st}$ 2019. After the glitch fix, shown score is the same as assessed score. The scores are similarly distributed among screener types. 

\begin{figure}[h]
\centering
\includegraphics[scale=0.25]{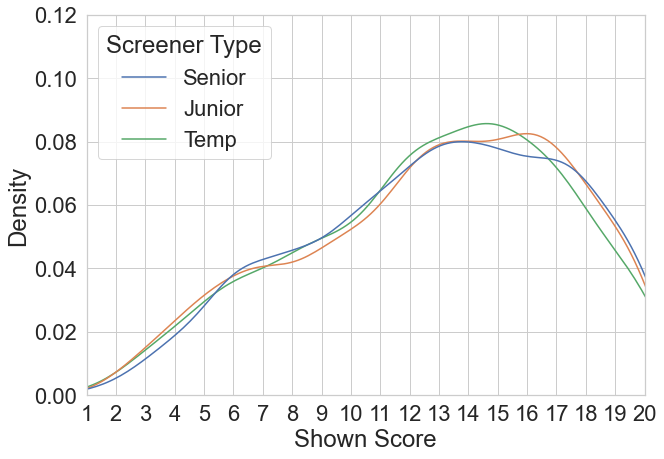}
\caption{Shown Scores Distribution By Screener Type. Area under each screener type is normalized to be 1. Shown scores are similarly distributed among screener types.}
\end{figure}

\subsection{Referral Characteristics}
\label{section:appendix_summary_graphs_referral_char}
\begin{figure}[h]
  \begin{minipage}[t]{0.48\textwidth}
   \centering
    \includegraphics[scale=0.2]{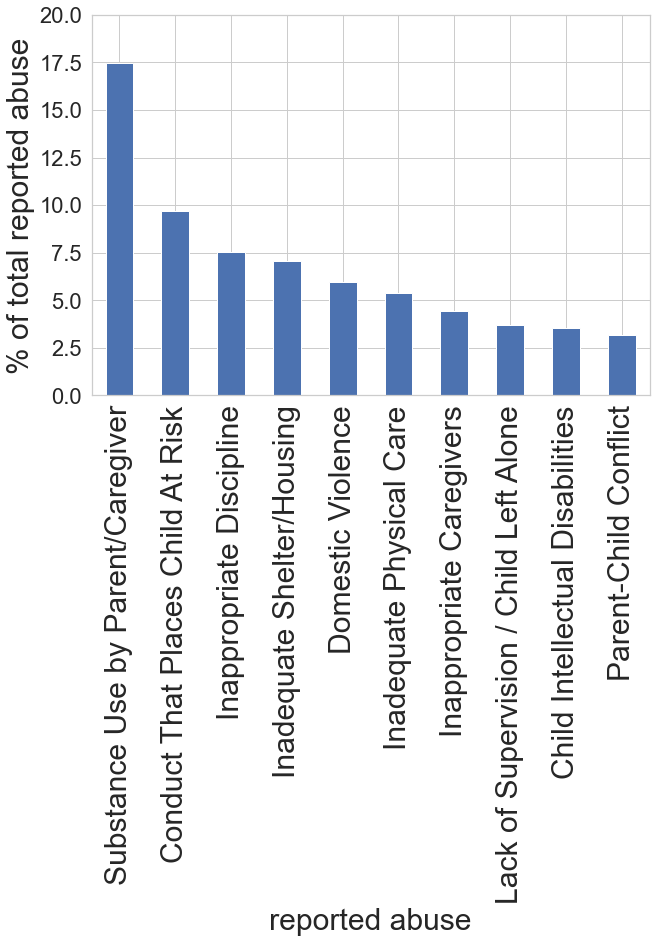}
    \\
    \subcaption{Top Ten Allegations for White Children}
  \end{minipage}
  \hfill
  \begin{minipage}[t]{0.48\textwidth}
    \includegraphics[scale=0.2]{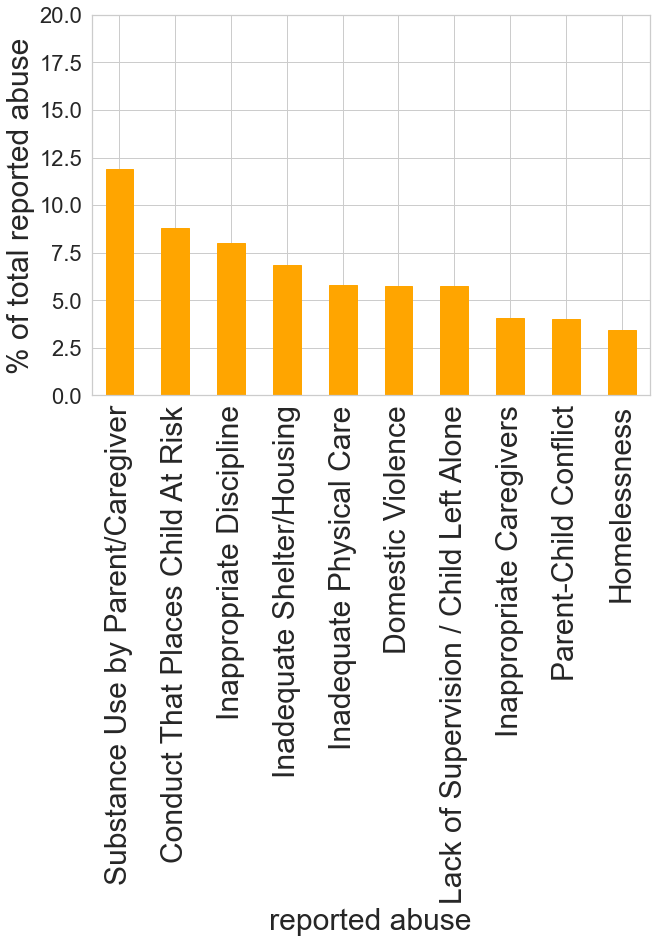}
    \\
     \subcaption{Top Ten Allegations for Black Children}
  \end{minipage}
 \caption{Top Ten Allegations by Race. Referrals associated with white and black children share common alleged abuse types. There is a higher percentage of white children being referred for alleged substance abuse by caregivers. However, child intellectual disabilities is a unique category in the top ten most common alleged abuse type for white children, as homelessness is for black children.}
 \label{fig:allegation-race}
\end{figure}

Referrals associated with white and black children differ in many aspects. We present summary statistics on the top 10 reported abuse types and their corresponding screen-in rates by race in Figure ~\ref{fig:allegation-race}. Referrals associated with white and black children share common alleged abuse types such as substance abuse by parent/caregiver, conduct that places child at risk, and so on. There is a higher percentage of white children being referred for alleged substance abuse by caregivers. However, child intellectual disabilities is a unique category in the top ten most common alleged abuse type for white children, as homelessness is for black children.

\begin{figure}[h]
  \begin{minipage}[t]{0.48\textwidth}
   \centering
    \includegraphics[scale=0.16]{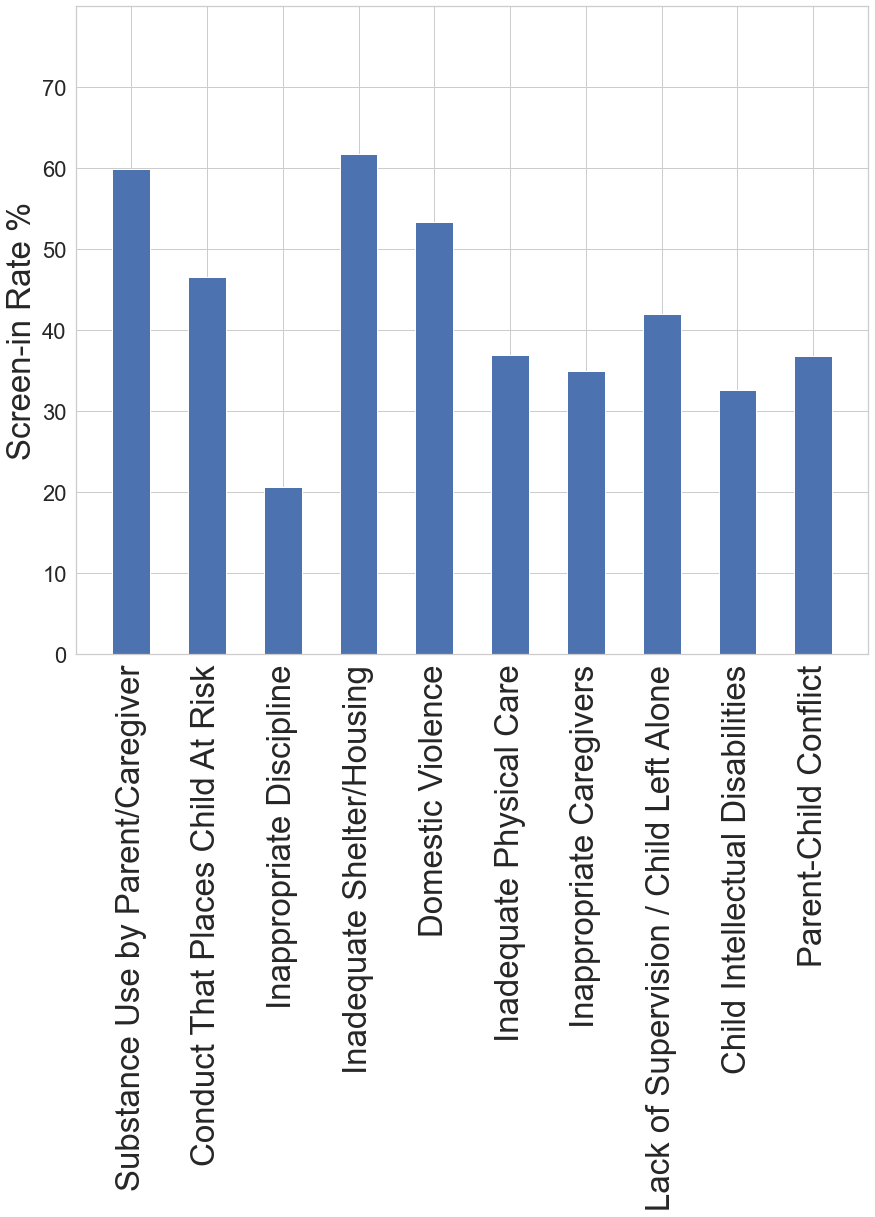}
    \\
    \subcaption{Screen-in Rates of the Top Ten Allegations for White Children}
  \end{minipage}
  \hfill
  \begin{minipage}[t]{0.48\textwidth}
   \centering
    \includegraphics[scale=0.16]{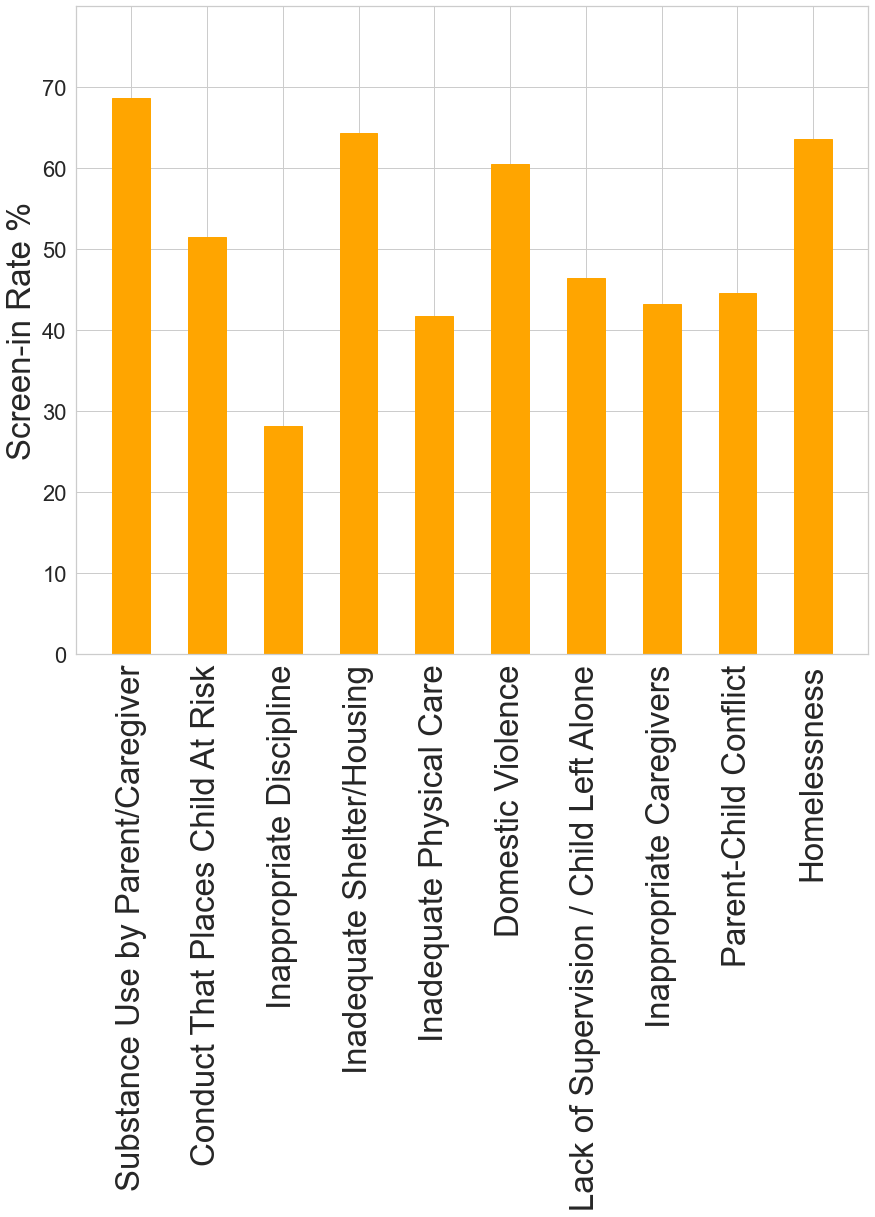}
        \\
     \subcaption{Screen-in Rates of the Top Ten Allegations for Black Children}
  \end{minipage}
 \caption{Screen-in Rate for Top Ten Allegations by Race. Screen-in rate varies for the same type of allegation between races. Screen-in rate is 10\% higher for black children than white for alleged substance abuse by parent/caregiver. Homelessness referrals are screened in at a high rate of above 60\% for black children.}
 \label{fig:screen-in-allegation-race}
\end{figure}

However, the screen-in rate varies for the same type of allegation between races. As shown in Figure ~\ref{fig:screen-in-allegation-race}, for the most common allegation, substance abuse by parent/caregiver, referrals on black children are about 10\% more likely to be screened in than white. Homelessness referrals are also screened in at a high rate of above 60\% for black children.

\section{Research Methodology}
\label{section:appendix_full_methodology}
We use short-hand in the main texts to show the equations in a more intuitive manner. The full equations and explanations are below. $\theta_1$ is a vector of estimated coefficients on each referral characteristics which we do not present in the tables:

\subsection{RQ1}
The following equation corresponds to equation ~\ref{eq:1}:
{\begin{align*}   
    & Y_{i,t,j} = \beta_0 + \beta_1 \tilde{S}_i + \beta_2 senior_{i,t} + \beta_3 junor_{i,t} + \beta_4 temp_{i,t} 
    \\ & +  \beta_5 senior_{i,t} * \tilde{S}_{i,t} + \beta_6 junior_{i,t} * \tilde{S}_{i,t} + \beta_7 temp_{i,t} * \tilde{S}_{i,t} 
    \\& + \theta_1 X_{i,t} + \gamma_t + \omega_j + \epsilon_{i,t,j}
\end{align*}}
$\beta{2}$ and $\beta{5}$ are subsequently dropped in estimation because senior screeners are used as the reference category.

\subsection{RQ2}
The following equation correspond to equation ~\ref{eq:2}
{\begin{align*}
& Y_{i,t,j} = \beta_0 + \beta_1 \tilde{S}_{i,t} + \beta_2 overestimation_{i,t} + \beta_3 correct_{i,t} + \beta_4 underestimation_{i,t} 
\\ & + \theta_1 X_{i,t} + \gamma_t + \omega_j + \epsilon_{i,t,j}
\end{align*}}
coefficients $\beta_2$ is dropped during estimation as the reference category. \\

The following equation corresponds to equation ~\ref{eq:3}
{\begin{align*}
& Y_{i,t} = \beta_0 + \beta_1 \tilde{S}_{i,t} + + \beta_2 senior_{i,t} + \beta_3 junior_{i,t} + \beta_4 temp_{i,t} 
\\ & + \beta_5 senior_{i,t} * overestimation_{i,t} + \beta_6 senior_{i,t} * correct_{i,t} + \beta_7 senior_{i,t} * underestimation_{i,t} 
\\ & + \beta_8 junior_{i,t} * overestimation_{i,t} + \beta_9 junior_{i,t} * correct_{i,t} + \beta_{10} junior_{i,t} * underestimation_{i,t} 
\\ & + \beta_{11} temp_{i,t} * overestimation_{i,t} + \beta_{12} temp_{i,t} * correct_{i,t} + \beta_{13} temp_{i,t} * underestimation_{i,t} 
\\ & + \theta_1 X_{i,t} + \gamma_t + \epsilon_{i,t}
\nonumber 
\end{align*}}
coefficients $\beta_2$, $\beta_5$, $\beta_6$, $\beta_7$ are dropped during estimation because senior screeners are the reference category. 

\subsection{RQ3}
The following equation corresponds to equation ~\ref{eq:4}:
\begin{align*}
& Y_{i,t,j} = \beta_0 + \beta_1 \tilde{S}_{i,t
} + \beta_2 Black_{i,t}  + \beta_3 overestimation_{i,t} + \beta_4 correct_{i,t} + \beta_5 underestimation_{i,t} \\ & + \theta_1 X_{i,t} + \gamma_t + \omega_j + \epsilon_{i,t,j}
\end{align*}

$\beta_4$ is dropped during estimation because correct estimation is used as the reference category. \\

The following equation corresponds to equation ~\ref{eq:5}:
\begin{align*}
& Y_{i,t,j} = \beta_0 + \beta_1 \tilde{S}_{i,t
} + \beta_2 senior_{i,t} + \beta_3 junior_{i,t} + \beta_4 temp_{i,t} \\ 
& + \beta_5 overestimation_{i,t} + \beta_6 correct_{i,t} + \beta_7 underestimation_{i,t} \\ 
& + \beta_8 Black * senior_{i,t} + \beta_9 Black * junior_{i,t} + \beta_{10} Black * temp_{i,t} + \theta_1 X_{i,t} + \gamma_t + \omega_j + \epsilon_{i,t,j}
\end{align*}

$\beta_6$ is dropped during estimation because correct estimation is used as the reference category. Referral associated with white children are used as the reference category for comparison against $\beta_8$,  $\beta_9$,  $\beta_10$. \\

The following equation corresponds to equation ~\ref{eq:6}:
\begin{align*}
& Y_{i,t,j} = \beta_0 + \beta_1 \tilde{S}_{i,t
} + \beta_2 senior_{i,t} + \beta_3 junior_{i,t} + \beta_4 temp_{i,t} 
\\ & + \beta_5 Black * senior_{i,t} + \beta_6 Black * junior_{i,t} + \beta_7 Black * temp_{i,t} + \theta_1 X_{i,t} + \gamma_t + \omega_j + \epsilon_{i,t,j}
\end{align*}

$\beta_2$ is dropped as the reference category. Referral associated with white children are used as the reference category for comparison against $\beta_5$,  $\beta_6$,  $\beta_7$. 

\section{Tables}
\label{section:appendix_tables}
We present the disaggregated results for RQ3. 
\begin{table}[h]
\caption{ESTIMATED EFFECTS OF RACE AND DISPLAY DISCREPANCY ON SCREEN-IN BY SCREENER EXPERIENCE}
\centering
\scalebox{0.6}
{
\def\sym#1{\ifmmode^{#1}\else\(^{#1}\)\fi}
\begin{tabular}{l*{6}{c}}
\hline\hline
&\multicolumn{1}{c}{(1)}&\multicolumn{1}{c}{(2)}&\multicolumn{1}{c}{(3)}&\multicolumn{1}{c}{(4)}&\multicolumn{1}{c}{(5)}&\multicolumn{1}{c}{(6)}\\
\hline
Shown Score &    0.0155\sym{***}&      0.0260\sym{***}&      0.0099\sym{***}&      0.0165\sym{***}&      0.0058\sym{***}&      0.0122\sym{***}\\
            &    (0.0009)         &    (0.0009)         &    (0.0008)         &    (0.0009)         &    (0.0009)         &    (0.0010)         \\
[1em]
Junior &       -0.1617\sym{***}&     -0.1502\sym{***}&     -0.0657\sym{*}  &     -0.0633\sym{*}  &     -0.0773\sym{*}  &     -0.0740\sym{**} \\
            &    (0.0304)         &    (0.0287)         &    (0.0306)         &    (0.0297)         &    (0.0291)         &    (0.0277)         \\
[1em]
Temp &    -0.1071\sym{**} &     -0.1017\sym{**} &     -0.0087         &     -0.0100         &     -0.0100         &     -0.0105         \\
            &    (0.0328)         &    (0.0302)         &    (0.0187)         &    (0.0179)         &    (0.0150)         &    (0.0146)         \\
[1em]
Black * Senior &       0.0297         &     -0.0049         &      0.0228\sym{*}  &      0.0021         &      0.0070         &     -0.0078         \\
            &    (0.0150)         &    (0.0138)         &    (0.0090)         &    (0.0089)         &    (0.0095)         &    (0.0091)         \\
[1em]
Black * Junior &      0.0409\sym{**} &      0.0098         &      0.0403\sym{*}  &      0.0214         &      0.0352         &      0.0217         \\
            &    (0.0121)         &    (0.0126)         &    (0.0174)         &    (0.0174)         &    (0.0193)         &    (0.0183)         \\
[1em]
Black * Temp &     0.0402\sym{**} &      0.0094         &      0.0163\sym{*}  &     -0.0013         &      0.0082         &     -0.0053         \\
            &    (0.0129)         &    (0.0116)         &    (0.0076)         &    (0.0076)         &    (0.0084)         &    (0.0085)         \\
[1em]
Month FE &  & &  Y   &  Y  & Y & Y\\
[1em]
Supervisor FE &  & &  Y   &  Y & Y & Y \\
[1em]
Referral Characteristics (No Race)  &  & &  &     &  Y   &  Y    \\
[1em]
Glitch Type &  & Y &    &  Y  & & Y\\
\hline
\(N\)       &       12,019         &       12,019         &       12,019        &       12,019         &      12,019      &      12,019         \\
\hline\hline
\multicolumn{7}{l}{\footnotesize Referrals on black and white children between 12/1/16-31/7/18 where both shown and assessed scores are available are included}\\
\multicolumn{7}{l}{\footnotesize The dependent variable is a binary variable, 1 if screened in and 0 otherwise.}\\
\multicolumn{7}{l}{\footnotesize The mean of the dependent variable is 0.4350, meaning the average screen in rate is 43.50.}\\
\multicolumn{7}{l}{\footnotesize Robust standard errors clustered by screener in parentheses. }\\
\multicolumn{7}{l}{\footnotesize \sym{*} \(p<0.05\), \sym{**} \(p<0.01\), \sym{***} \(p<0.001\)}\\
\end{tabular}
}
\label{table:race_discrepancy_by_screener_glitch}
\end{table}
\begin{table}[H]
\caption{ESTIMATED EFFECTS OF RACE ON SCREEN-IN POST GLITCH BY SCREENER EXPERIENCE}
\centering
\scalebox{0.6}
{
\def\sym#1{\ifmmode^{#1}\else\(^{#1}\)\fi}
\begin{tabular}{lccc}
\hline\hline
            &\multicolumn{1}{c}{(1)}&\multicolumn{1}{c}{(2)}&\multicolumn{1}{c}{(3)}\\
\hline
Shown Score &     0.0282\sym{***}&      0.0237\sym{***}&      0.0188\sym{***}\\
            &    (0.0013)         &    (0.0011)         &    (0.0013)         \\
[1em]
Junior &    -0.0869\sym{**} &     -0.0327         &     -0.0616\sym{**} \\
            &    (0.0304)         &    (0.0253)         &    (0.0212)         \\
[1em]
Temp &      -0.1001\sym{***}&     -0.0427         &     -0.0731\sym{**} \\
            &    (0.0230)         &    (0.0246)         &    (0.0245)         \\
[1em]
Black * Senior &      0.0330         &      0.0256         &     -0.0163         \\
            &    (0.0205)         &    (0.0171)         &    (0.0162)         \\
[1em]
Black * Junior &      0.0660\sym{***}&      0.0474\sym{***}&      0.0159         \\
            &    (0.0111)         &    (0.0118)         &    (0.0085)         \\
[1em]
Black * Temp &     0.0706         &      0.0538         &      0.0267         \\
            &    (0.0433)         &    (0.0366)         &    (0.0335)         \\
[1em]
Month FE & &  Y & Y \\
[1em]
Supervisor FE & &  Y & Y  \\
[1em]
Referral Characteristics (No Race)  & &   & Y
\\
\hline
\(N\)       &        7,351         &        7,351         &        7,351         \\
\hline\hline
\multicolumn{4}{l}{\footnotesize Referrals on black and white children between 2/1/19-12/31/19 are included}\\
\multicolumn{4}{l}{\footnotesize The dependent variable is a binary variable, 1 if screened in and 0 otherwise.}\\
\multicolumn{4}{l}{\footnotesize The mean of the dependent variable is 0.4391 , meaning the average screen in rate is 43.91.}\\
\multicolumn{4}{l}{\footnotesize Robust standard errors clustered by screener in parentheses. }\\
\multicolumn{4}{l}{\footnotesize \sym{*} \(p<0.05\), \sym{**} \(p<0.01\), \sym{***} \(p<0.001\)}\\
\end{tabular}
}
\label{table:race_discrepancy_by_screener_post_glitch}
\end{table}

\section{Alternative Analyses using Logistic Regression on Main Results}
\label{section:appendix_logit}
We present alternative specifications using logistic regression for our main results. Tables ~\ref{table:utilization_of_scores_by_screener_logit}
~\ref{table:non_madatory_screen_in_by_screener_logit)}~\ref{table:race_discrepancy_on_screen_in_logit}
~\ref{table:race_discrepancy_on_screen_in_post_glitch_logit} correspond to Tables ~\ref{table:utilization_of_scores_by_screener}
~\ref{table:non_madatory_screen_in_by_screener}
~\ref{table:race_discrepancy_on_screen_in}
~\ref{table:race_discrepancy_on_screen_in_post_glitch} accordingly. When considering fixed effects, logistic regression drops observations where there are insufficient variations by supervisor fixed effects. We observe persistent statistical significance of the results in the alternative model as in the linear models. We bold the coefficients for which we made statements about in the main paper for comparison. 
\begin{table}[h]
\caption{LR: SCREENERS' UTILIZATION OF RISK SCORES}
\centering
\scalebox{0.6}
{
\def\sym#1{\ifmmode^{#1}\else\(^{#1}\)\fi}
\begin{tabular}{l*{4}{c}}
\hline\hline
            &\multicolumn{1}{c}{(1)}&\multicolumn{1}{c}{(2)}&\multicolumn{1}{c}{(3)}&\multicolumn{1}{c}{(4)}\\
\hline
Shown Score &     0.0755\sym{***}&      0.0762\sym{***}&      \textbf{0.0478}\sym{***}&      0.0478\sym{***}\\
            &    (0.0048)         &    (0.0071)         &    (0.0058)         &    (0.0057)         \\
[1em]
Junior &     -0.8401\sym{***}&     -0.5967\sym{***}&     \textbf{-0.8498}\sym{***}&     -0.8421\sym{***}\\
            &    (0.1532)         &    (0.1756)         &    (0.1672)         &    (0.1661)         \\
[1em]
Temp &   -0.6015\sym{***}&     -0.3236         &     \textbf{-0.3615}\sym{*}  &     -0.3515\sym{*}  \\
            &    (0.1768)         &    (0.1813)         &    (0.1419)         &    (0.1411)         \\
[1em]
Junior * Shown Score &     0.0246\sym{**} &      0.0250\sym{**} &      \textbf{0.0314}\sym{**} &      0.0310\sym{**} \\
            &    (0.0077)         &    (0.0095)         &    (0.0097)         &    (0.0097)         \\
[1em]
Temp * Shown Score &     0.0077         &      0.0055         &      \textbf{0.0052}         &      0.0046         \\
            &    (0.0074)         &    (0.0089)         &    (0.0082)         &    (0.0081)         \\
[1em]
Month FE &   &  Y   &  Y  & Y\\
[1em]
Supervisor FE &   &  Y   &  Y & Y \\
[1em]
Referral Characteristics  &   &     &  Y   &      \\
[1em]
Referral Characteristics (No Race) &   &     &  & Y\\
\hline
\(N\)       &       28,736         &       28,693         &       28,632         &       28,632        \\
\hline\hline
\multicolumn{5}{l}{\footnotesize Referrals of all races are included.}\\
\multicolumn{5}{l}{\footnotesize Some observations are dropped due to insufficient variation by supervisor fixed effects}\\
\multicolumn{5}{l}{\footnotesize Robust standard errors clustered by screener in parentheses. }\\
\multicolumn{5}{l}{\footnotesize \sym{*} \(p<0.05\), \sym{**} \(p<0.01\), \sym{***} \(p<0.001\)}\\
\end{tabular}
}
\label{table:utilization_of_scores_by_screener_logit}
\end{table}
\begin{table}[h]
\caption{LR: SCREEN-IN OF NON-MANDATORY REFERRALS BY SCREENER AND DISPLAY DISCREPANCY}
\centering
\scalebox{0.6}
{
\def\sym#1{\ifmmode^{#1}\else\(^{#1}\)\fi}
\begin{tabular}{l*{4}{c}}
\hline\hline
            &\multicolumn{1}{c}{(1)}&\multicolumn{1}{c}{(2)}&\multicolumn{1}{c}{(3)}&\multicolumn{1}{c}{(4)}\\
\hline
Shown Score &       0.1084\sym{***}&      0.1086\sym{***}&      0.0781\sym{***}&      0.0795\sym{***}\\
            &    (0.0051)         &    (0.0070)         &    (0.0075)         &    (0.0074)         \\
[1em]
Junior &      -0.8252\sym{***}&     -0.5964\sym{**} &     -0.7801\sym{***}&     -0.7817\sym{***}\\
            &    (0.1306)         &    (0.1924)         &    (0.2124)         &    (0.2188)         \\
[1em]
Temp &     -0.6365\sym{***}&     -0.2246         &     -0.3149         &     -0.3141         \\
            &    (0.1542)         &    (0.1746)         &    (0.1729)         &    (0.1748)         \\
[1em]
Senior * Overestimation &     -0.0141         &      0.0218         &     \textbf{-0.2011}         &     -0.1959         \\
            &    (0.0617)         &    (0.1024)         &    (0.1410)         &    (0.1407)         \\
[1em]
Junior * Overestimation &      0.2909         &      0.3488         &      \textbf{0.2381}         &      0.2475         \\
            &    (0.1912)         &    (0.2404)         &    (0.2306)         &    (0.2310)         \\
[1em]
Temp * Overestimation &     0.1156         &      0.1333         &  \textbf{0.0030}         &      0.0029         \\
            &    (0.0758)         &    (0.0914)         &    (0.0993)         &    (0.0975)         \\
[1em]
Senior * Underestimation &       1.2389\sym{***}&      1.3126\sym{***}&      \textbf{1.1101}\sym{***}&      1.1110\sym{***}\\
            &    (0.0624)         &    (0.1116)         &    (0.1260)         &    (0.1294)         \\
[1em]
Junior * Underestimation &       1.4036\sym{***}&      1.5492\sym{***}&      \textbf{1.3612}\sym{***}&      1.3492\sym{***}\\
            &    (0.1588)         &    (0.2056)         &    (0.1674)         &    (0.1716)         \\
[1em]
Temp * Underestimation &       1.4950\sym{***}&      1.4449\sym{***}&      \textbf{1.2846}\sym{***}&      1.2835\sym{***}\\
            &    (0.0792)         &    (0.1086)         &    (0.1286)         &    (0.1291)         \\
[1em]
Month FE &   &  Y   &  Y  & Y\\
[1em]
Supervisor FE &   &  Y   &  Y & Y \\
[1em]
Referral Characteristics  &   &     &  Y   &      \\
[1em]
Referral Characteristics (No Race) &   &     &  & Y\\
\hline
\(N\)       &       12,981         &       12,936         &       12,874         &       12,874         \\
\hline\hline
\multicolumn{5}{l}{\footnotesize Referrals of all races between 12/1/16-31/7/18 where both shown and assessed scores are available}\\
\multicolumn{5}{l}{\footnotesize and no mandatory screen-in required are included.}\\
\multicolumn{5}{l}{\footnotesize Some observations are dropped due to insufficient variation by supervisor fixed effects}\\
\multicolumn{5}{l}{\footnotesize Robust standard errors clustered by screener in parentheses. }\\
\multicolumn{5}{l}{\footnotesize \sym{*} \(p<0.05\), \sym{**} \(p<0.01\), \sym{***} \(p<0.001\)}\\
\end{tabular}
}
\label{table:non_madatory_screen_in_by_screener_logit)}
\end{table}
\begin{table}[h]
\caption{LR: ESTIMATED EFFECTS OF RACE AND DISPLAY DISCREPANCY ON SCREEN-IN}
\centering
\scalebox{0.6}
{
\def\sym#1{\ifmmode^{#1}\else\(^{#1}\)\fi}
\begin{tabular}{l*{6}{c}}
\hline\hline
            &\multicolumn{1}{c}{(1)}&\multicolumn{1}{c}{(2)}&\multicolumn{1}{c}{(3)}&\multicolumn{1}{c}{(4)}&\multicolumn{1}{c}{(5)}&\multicolumn{1}{c}{(6)}\\
\hline
Shown Score &     0.0656\sym{***}&      0.1192\sym{***}&      0.0728\sym{***}&      0.1271\sym{***}&      0.0458\sym{***}&      0.1027\sym{***}\\
            &    (0.0046)         &    (0.0051)         &    (0.0072)         &    (0.0081)         &    (0.0075)         &    (0.0093)         \\
[1em]
Black       &      \textbf{0.1473}\sym{***}&      \textbf{0.0083}         &      \textbf{0.1765}\sym{***}&      \textbf{0.0332}         &      \textbf{0.1004}         &     \textbf{-0.0163}         \\
            &    (0.0368)         &    (0.0369)         &    (0.0441)         &    (0.0482)         &    (0.0603)         &    (0.0611)         \\
[1em]
Overestimation &                  &     -0.1007\sym{*}  &                     &     -0.1327         &                     &     -0.3130\sym{***}\\
            &                     &    (0.0467)         &                     &    (0.0729)         &                     &    (0.0834)         \\
[1em]
Underestimation &        &      1.2093\sym{***}&                     &      1.1995\sym{***}&                     &      1.0446\sym{***}\\
            &                     &    (0.0551)         &                     &    (0.0731)         &                     &    (0.0864)         \\
[1em]
Month FE &  & &  Y   &  Y  & Y & Y\\
[1em]
Supervisor FE &  & &  Y   &  Y & Y & Y \\
[1em]
Referral Characteristics (No Race)  &  & &  &     &  Y   &  Y    \\
\hline
\(N\)       &       12,019         &       12,019         &       11,981        &       11,981         &       11,922         &       11,922        \\
\hline\hline
\multicolumn{7}{l}{\footnotesize Referrals on black and white children between 12/1/16-31/7/18 where both shown and assessed scores are available are included}\\
\multicolumn{7}{l}{\footnotesize Some observations are dropped due to insufficient variation by supervisor fixed effects}\\
\multicolumn{7}{l}{\footnotesize Robust standard errors clustered by screener in parentheses. }\\
\multicolumn{7}{l}{\footnotesize \sym{*} \(p<0.05\), \sym{**} \(p<0.01\), \sym{***} \(p<0.001\)}\\
\end{tabular}
}
\label{table:race_discrepancy_on_screen_in_logit}
\end{table}

\begin{table}[h]
\caption{LR: ESTIMATED EFFECTS OF RACE ON SCREEN-IN POST GLITCH}
\centering
\scalebox{0.6}
{
\def\sym#1{\ifmmode^{#1}\else\(^{#1}\)\fi}
\begin{tabular}{lccc}
\hline\hline
            &\multicolumn{1}{c}{(1)}&\multicolumn{1}{c}{(2)}&\multicolumn{1}{c}{(3)}\\
\hline
Shown Score &       0.1237\sym{***}&      0.1262\sym{***}&      0.1172\sym{***}\\
            &    (0.0063)         &    (0.0056)         &    (0.0093)         \\
[1em]
Black       &     0.2416\sym{***}&      0.2199\sym{***}&      \textbf{0.0502}         \\
            &    (0.0521)         &    (0.0557)         &    (0.0648)         \\
[1em]
Month FE & &  Y & Y \\
[1em]
Supervisor FE & &  Y & Y  \\
[1em]
Referral Characteristics (No Race)  & &   & Y \\
\hline
\(N\)       &        7,351         &       7,278         &        7,253         \\
\hline\hline
\multicolumn{4}{l}{\footnotesize Referrals on black and white children between 2/1/19-12/31/19 are included}\\
\multicolumn{4}{l}{\footnotesize Some observations are dropped due to insufficient variation by supervisor fixed effects}\\
\multicolumn{4}{l}{\footnotesize Robust standard errors clustered by screener in parentheses. }\\
\multicolumn{4}{l}{\footnotesize \sym{*} \(p<0.05\), \sym{**} \(p<0.01\), \sym{***} \(p<0.001\)}\\
\end{tabular}
}
\label{table:race_discrepancy_on_screen_in_post_glitch_logit}
\end{table}

\end{document}